%% file: main.tex
\documentclass[%
 reprint,
superscriptaddress,
 amsmath,amssymb,
 aps,
prapplied,
]{revtex4-2}

\usepackage{graphicx}
\usepackage{dcolumn}
\usepackage{bm}
\usepackage{physics}
\usepackage{amsmath,amssymb}
\usepackage{subfiles}
\usepackage{xcolor}

\newcommand{\textapprox}{\raisebox{0.5ex}{\texttildelow}}

\begin{document}


\title{Incorporating multi-qubit exchange coupling effects between transmon qubits in Maxwell-Schr\"{o}dinger numerical methods}

\author{Ghazi Khan}
\affiliation{%
Elmore Family School of Electrical and Computer Engineering, Purdue University, Indiana 47906, USA
}%
\author{Thomas E. Roth}%
\email{Contact author: rothte@purdue.edu}
\affiliation{%
Elmore Family School of Electrical and Computer Engineering, Purdue University, Indiana 47906, USA
}%
\affiliation{Purdue Quantum Science and Engineering Institute, Purdue University, Indiana 47907, USA}




\date{\today}

\input{Abstract}
\maketitle
\input{Introduction}
\input{Background}

\input{Two-Qubit-Quantum}

\input{Numerical_Results}
\input{Conclusion}

\appendix
\input{Appendix_SQ}

\input{Appendix_SW}
\input{Open_Quantum_System}

\bibliography{main}

\end{document}

%% file: Abstract.tex
\begin{abstract}
Superconducting qubits have emerged as one of the leading platforms for realizing practical quantum computers. Accurate modeling of these devices is essential for predicting performance, improving design, and optimizing control. Many existing modeling approaches currently rely on lumped circuit approximations or other simplified treatments that can be limited in resolving the interplay between the qubit dynamics and the underlying electromagnetic circuitry, leading to significant experimental deviations from numerical predictions at times. To address many of these limitations, methods that self-consistently solve the Schr\"{o}dinger equation for qubit dynamics with the classical Maxwell's equations (or their transmission line variant) have been developed and shown to accurately predict a wide range of effects related to superconducting qubit control and readout. Despite the successes of these Maxwell-Schr\"{o}dinger methods, they have not been able to consider multi-qubit effects that give rise to qubit-qubit entanglement to date. Here, we address this limitation by rigorously deriving how multi-qubit exchange coupling effects between transmon qubits can be embedded into Maxwell-Schr\"{o}dinger numerical methods. To support this, we build on earlier first-principles derivations of Maxwell-Schr\"{o}dinger methods for the specific case of two transmon qubits coupled together through a common electromagnetic system in the dispersive regime of circuit quantum electrodynamics. To aid in validating aspects of the Maxwell-Schr\"{o}dinger framework, we also provide a new interpretation of Maxwell-Schr\"{o}dinger methods as an efficient simulation strategy to capture the class of non-Markovian open quantum system dynamics when the Born approximation holds and the electromagnetic system is in a coherent state. Our numerical results demonstrate that these effects can give rise to strong classical crosstalk that can significantly alter multi-qubit dynamics, which we demonstrate here for the cross-resonance gate. These strong classical crosstalk effects have been noted in cross-resonance experiments, but previous quantum theory and device analysis could not explain their origin. Our Maxwell-Schr\"{o}dinger methods provide a way to predictively model these effects, and can be used in the future to create new device and control pulse designs that mitigate these deleterious effects.
\end{abstract}

%% file: Introduction.tex
\section{Introduction}
Superconducting qubits are a leading hardware platform for many quantum information technologies \cite{krantz2019quantum,gu2017microwave,blais2021circuit}, with a particular emphasis in the area of quantum computing. This circuit quantum electrodynamics (cQED) platform has been used for various experimental demonstrations of quantum error correction \cite{zhao2022realization,yang2024coupler,google2025quantum}, as well as for a testbed of quantum simulations \cite{meng2024simulating,rosen2024synthetic,rosenberg2024dynamics,altman2021quantum}. However, these devices have also found use in other areas, such as exploring key fundamental physics ideas \cite{dixit2021searching,braggio2025quantum,zhang2025experimental,wang2025probing}. To support these myriad uses, constant advancements are being made at every level to improve coherence times \cite{somoroff2023millisecond,tuokkola2025methods,bland2025millisecond}, achieve higher-fidelity gates \cite{li2024realization,lin202524}, and push towards better scalability \cite{spring2025fast,song2025realization,google2025quantum}. While many directions continue to exist to improve device performance, as these systems mature an increasing focus is starting to be placed on how we numerically model them to improve the success of the design process and enable systematic engineering through simulation \cite{mohseni2024build,Levenson-Falk_2025,shanto2024squadds,elkin2025opportunities}.

Of particular interest are general-purpose numerical methods; i.e., ones that can be applied in a uniform way to an arbitrary device design or layout without requiring the numerical method or solution procedure to be reformulated. Early efforts in this direction were black-box quantization approaches \cite{nigg2012black,solgun2014blackbox}, which used electromagnetic (EM) simulations of a linearized physical model to characterize the Hamiltonian of weakly-anharmonic qubits, such as transmons \cite{koch2007charge,roth2022transmon}, by treating the nonlinearity of the qubits perturbatively. The EM simulations performed in this approach were ``driven simulations'' that included ports in place of the Josephson junctions in the qubits to extract multi-port impedance parameters. More recently, the energy participation ratio (EPR) quantization approach was introduced to avoid some of the inconvenient processing aspects of black-box quantization approaches by performing EM eigenmode simulations instead of driven simulations \cite{minev2021energy}. Despite the implementation differences, this approach still treats the nonlinearity of the qubits perturbatively. Unfortunately, this perturbative treatment results in slow numerical convergence in terms of the number of Fock states per mode that must be used in constructing the Hamiltonian for both the black-box and EPR quantization approaches to achieve accurate results \cite{moon2024analytical}. 

This slow convergence can be circumvented if one leverages the prior knowledge about the qubits that are being designed in the device  (e.g., that they are transmons) to treat the nonlinearity directly, as was done in \cite{moon2024analytical} to achieve a \textapprox280x speedup over EPR in simulating a parametric frequency sweep of a two-qubit device with three EM modes. While such speedups are appealing for many practical small-scale simulations, all of these methods still fundamentally suffer from the rapid exponential growth that occurs in generating a matrix representation of the system Hamiltonian \cite{moon2024computational}. This problem is further exacerbated by the need to often include multiple EM modes for each physical EM resonator to accurately capture effects like multi-qubit exchange coupling \cite{filipp2011multimode,khan2024field}. Although tensor network compression could be used to circumvent the exponential growth, the global couplings that can exist between EM modes and qubits could still be problematic for existing techniques \cite{ryu2023efficient}. Given that even extracting a large number of EM modes can be extremely challenging for cQED devices \cite{elkin2025opportunities}, alternative modeling strategies that are more computationally scalable are a significant need.

Generally, more scalable strategies require exploiting some aspects of the expected physics directly in the model formulation. For cQED devices, the dispersive operating regime means that most of the influence of quantum EM effects are through vacuum fluctuations. It is a common strategy in macroscopic quantum electrodynamics (QED) to relate the effects of EM vacuum fluctuations to EM dyadic Green's functions \cite{scheel2008macroscopic,moon2024computational}, which can be evaluated numerically much more efficiently than computing a large number of EM eigenmodes \cite{elkin2025opportunities}. Similar strategies have started to be exploited for cQED devices, where instead of EM dyadic Green's functions one can make use of impedance parameters (or other microwave network parameters) to directly compute properties like spontaneous emission rates or multi-qubit exchange coupling rates \cite{houck2008controlling,solgun2019simple,roth2022full,solgun2022direct,khan2024field,labarca2024toolbox}. The key benefit of these approaches is to avoid performing an EM eigenmode decomposition or needing to incorporate these EM modes explicitly in the system Hamiltonian.

While this is attractive from a modeling efficiency perspective, it is possible for such approaches to miss dynamical effects due to the interactions between the qubits and the surrounding EM circuitry. One efficient way to incorporate such effects is through a semiclassical Maxwell-Schr\"{o}dinger method that self-consistently solves the coupled classical Maxwell's equations with the Schr\"{o}dinger equation \cite{moon2024computational} (Maxwell-Bloch equations are also in this vein of approach \cite{scully1997quantum,jirauschek2019optoelectronic}). These approaches have shown in engineered systems that the back-action of the qubits on the applied EM fields can be significant for applications of light-matter interactions \cite{takeuchi2015maxwell,lorin2007numerical}, plasmonic nano-devices \cite{sukharev2023efficient}, quantum dots \cite{capua2012finite}, and superconducting qubits \cite{roth2024maxwell}, among others \cite{moon2024computational}. In the case of superconducting qubits, these methods have been able to accurately model dispersive readout and ac-Stark shifts for transmon and fluxonium qubits \cite{roth2024maxwell} and for devices with Purcell filters \cite{elkin2024epeps}, as well as predict the fidelity of multiplexed readout chains incorporating Josephson traveling-wave parametric amplifiers \cite{elkin2025ims,elkin2025finite}.

Despite these interesting use cases, Maxwell-Schr\"{o}dinger methods have not been able to rigorously model multi-qubit effects that give rise to entanglement between qubits, limiting their utility for optimizing qubit control. Here, we show how to rigorously incorporate such effects into Maxwell-Schr\"{o}dinger methods for transmon qubits operating in the dispersive regime where multi-qubit exchange coupling \cite{filipp2011multimode,khan2024field} is due to EM vacuum fluctuations. To support this, we build on an earlier first-principles derivation of Maxwell-Schr\"{o}dinger methods \cite{vsindelka2010derivation} to provide a new interpretation of them as an efficient way to model a class of non-Markovian open quantum system effects. Our results further demonstrate that these effects can give rise to a form of ``classical crosstalk'' that has been noted in experiments to significantly influence the multi-qubit dynamics of cross-resonance gates \cite{sheldon2016procedure}, but for which prior theory could not describe from a first-principles perspective \cite{magesan2020effective}. Beyond providing new insight into the origin of these deleterious effects, our multi-qubit Maxwell-Schr\"{o}dinger method also provides a predictive modeling capability to assist in designing new physical layouts and control pulses to mitigate their influence.

The remainder of this paper is organized as follows. In Section \ref{sec:maxwell-schrodinger-background}, we review the needed details about single-qubit Maxwell-Schr\"{o}dinger models. We then develop in Section \ref{sec:multi-qubit} a first-principles quantum formalism for multi-qubit transmon devices operating in the dispersive regime to determine how a multi-qubit Maxwell-Schr\"{o}dinger model can be implemented. Following this, we present numerical results in Section \ref{sec:results} of the multi-qubit Maxwell-Schr\"{o}dinger model, including results focusing on the ``classical crosstalk'' effect noted in prior cross-resonance gate experiments \cite{sheldon2016procedure}. Mathematical details about the derivations are included in Appendices \ref{app:single-qubit} and \ref{sec:appendix-SW}, while Appendix \ref{sec:Open-Quant} focuses on developing the open quantum system interpretation of Maxwell-Schr\"{o}dinger methods in the context of single-qubit control to support the formulation of the main text.  

%% file: Background.tex
\section{Maxwell-Schr\"{o}dinger Method Background}
\label{sec:maxwell-schrodinger-background}
\begin{figure}[t!]
    \centering
    \includegraphics[width=0.75\linewidth]{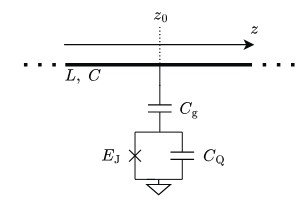}
    \caption{Schematic for deriving the Maxwell-Schr\"{o}dinger method for a transmon and transmission line system.} 
    \label{fig:maxwell-schrodinger-derivation}
\end{figure}

In this section, we briefly discuss the Maxwell-Schr\"{o}dinger method of \cite{roth2024maxwell} to assist with the discussions on how to incorporate multi-qubit effects into it, as well as the later open quantum system interpretations of this method. The basic setup for formulating the Maxwell-Schr\"{o}dinger method for superconducting qubits is shown in Fig. \ref{fig:maxwell-schrodinger-derivation}, where we specifically consider the case of a transmon qubit capacitively coupled to a transmission line. In \cite{roth2024maxwell}, a Hamiltonian mechanics derivation was considered to determine the coupled dynamical equations suitable for this setup. 

When representing the qubit wavefunction in the position-like phase basis of superconductors with phase variable $\varphi$, we get a Schr\"{o}dinger equation of
\begin{multline}
	\big[4 E_\mathrm{C} \big(\!-\!i\partial_\varphi  \big)^2  - E_\mathrm{J} \cos\varphi \big] \psi(\varphi,t) 
	\\ -i 2e\beta \big( \partial_t \phi(z_0,t) \big) \partial_\varphi \psi(\varphi,t)  = i\hbar\partial_t \psi(\varphi,t),
	\label{eq:schro-wave4}
\end{multline}
where $E_\mathrm{C}$ is the qubit charging energy, $E_\mathrm{J}$ is the Josephson energy, $\beta = C_\mathrm{g}/(C_\mathrm{g}+C_\mathrm{Q})$, and $\phi(z_0,t)$ is the transmission line node flux at the coupling point $z_0$. Similarly, the dynamical equation for the transmission line is given by the wave equation
\begin{align}
	\partial_z^2 \phi - LC\partial_t^2\phi = -\delta(z-z_0) L 2e\beta \partial_t \langle {n}(t)\rangle,
	\label{eq:phi-coupled-wave}
\end{align}
where $L$ and $C$ are the per-unit-length inductance and capacitance, respectively. Further, $\langle n(t) \rangle$ is the expectation value of the qubit charge operator $\hat{n}$, which in the phase basis is given by
\begin{align}
	\langle {n}(t) \rangle = \int_{-\pi}^\pi \psi^*(\varphi,t) \big(\!-\!i\partial_\varphi \psi(\varphi,t) \big) d\varphi.
\end{align}
The term on the right-hand side of (\ref{eq:phi-coupled-wave}) can be interpreted as a semiclassical current source that couples the qubit dynamics into the transmission line wave equation. Similarly, the $\partial_t \phi(z_0,t)$ term in (\ref{eq:schro-wave4}) couples the effective voltage from the transmission line onto the qubit to modify the qubit state. 

By self-consistently evolving (\ref{eq:schro-wave4}) and (\ref{eq:phi-coupled-wave}) in time, the back-action of the qubit dynamics on the EM circuitry (and vice-versa) can be efficiently modeled \cite{roth2024maxwell}. To accomplish this numerically, a leap-frog time-marching strategy was adopted in \cite{roth2024maxwell} similar to what is done in finite-difference time-domain simulations of Maxwell's equations \cite{teixeira2023finite}. This corresponds to discretizing and updating $\psi$ and $\phi$ on staggered temporal grids such that the quantities needed in (\ref{eq:schro-wave4}) related to $\partial_t\phi$ are known from previous updates to (\ref{eq:phi-coupled-wave}), and similar for updating (\ref{eq:phi-coupled-wave}) due to $\partial_t \langle {n}(t)\rangle$. To handle the spatial discretization of (\ref{eq:schro-wave4}) and (\ref{eq:phi-coupled-wave}), many different strategies could be followed, such as finite differences or the finite element method (FEM) \cite{jin2011theory}. Here, we discretize (\ref{eq:schro-wave4}) using the eigenstates of the free transmon Hamiltonian and use FEM to discretize (\ref{eq:phi-coupled-wave}). The FEM discretization of (\ref{eq:phi-coupled-wave}) is selected to make the method more readily extensible to perform 3D full-wave simulations \cite{elkin2025opportunities}. Importantly, by using a FEM discretization the resulting Maxwell-Schr\"{o}dinger method can also be applied easily to arbitrary circuit schematics without requiring any reformulations of the underlying numerical method.

This simulation strategy was demonstrated for various qubit control scenarios in \cite{roth2024maxwell}, and was also shown to quantitatively capture a wide range of dispersive regime effects relevant to qubit readout \cite{roth2024maxwell,elkin2024epeps,elkin2025ims}. However, quantitative validation of the method for qubit control scenarios was not demonstrated in these earlier works. In Appendix \ref{sec:Open-Quant}, we show this quantitative validation for qubit control scenarios. 

As part of this, we further demonstrate that Maxwell-Schr\"{o}dinger methods can be viewed as an efficient simulation strategy to capture certain non-Markovian open quantum system effects. Specifically, the Maxwell-Schr\"{o}dinger method corresponds to a Born approximation with the further approximation that the EM circuitry is in a coherent state when accounting for the influence of the EM circuitry on the qubit dynamics. While this represents a small class of non-Markovian quantum effects within the broader theory of open quantum systems, these effects can be prevalent for cQED devices due to the ways they are engineered. Further, as we will show in Section \ref{sec:results}, these non-Markovian open quantum system effects can have dramatic impacts on multi-qubit gates, such as cross-resonance gates. Such influences have been observed experimentally in \cite{sheldon2016procedure} and were attributed to a ``classical crosstalk'' term that was not able to be explained from the physical layout of the device or justified from the quantum theory in \cite{magesan2020effective}. Here, we show how a multi-qubit Maxwell-Schr\"{o}dinger method can predictively model these effects, providing a new resource for engineering devices to mitigate them.

%% file: Two-Qubit-Quantum.tex
\section{Multi-Qubit Effects in Maxwell-Schr\"{o}dinger Methods}
\label{sec:multi-qubit}
We now turn to extending the Maxwell-Schr\"{o}dinger framework by rigorously determining how to incorporate multi-qubit effects into it. In Section \ref{subsec:quantum-formulation}, we build on a first-principles derivation of the Maxwell-Schr\"{o}dinger framework \cite{vsindelka2010derivation} to consider the specific case of two transmon qubits coupled to each other through a common EM system in the dispersive regime. We then discuss in Section \ref{subsec:multi-qubit-Maxwell-Schrodinger} what changes are required to the Maxwell-Schr\"{o}dinger equations of motion to capture the multi-qubit effects derived in Section \ref{subsec:quantum-formulation}.

\subsection{Quantum Formulation}
\label{subsec:quantum-formulation}
To begin, we note that the Hamiltonian for two transmon qubits coupled to each other through a common EM system can be given by \cite{khan2024field}
\begin{multline}
    \label{EQ:Two-Qubit-Ham}
    \hat{H}_{2}=\hbar\sum_{j}\sum_{l=1}^{2}q_{j}^{(l)}\ket{j}^{(l)}\bra{j}^{(l)}+\hbar\sum_{k}\omega_k\hat{a}_k^\dag \hat{a}_k\\
+\sum_{j,k}\sum_{l=1}^2\hbar\biggl[g_{j,k}^{(l)}\ket{j}^{(l)}\bra{j+1}^{(l)}\hat{a}_k^\dag+\textnormal{H.c.}\biggr].
\end{multline}
Here, $q_j^{(l)}$ corresponds to the eigenfrequency of the $j$-th eigenstate $\ket{j}^{(l)}$ of the $l$-th free transmon Hamiltonian and $\omega_k$ is the frequency associated with the $k$-th mode of the EM system with corresponding annihilation (creation) operator $\hat{a}_k$ ($\hat{a}_k^\dag$). Further, the coupling between the $j$-th level of the $l$-th transmon with the $k$-th EM mode is $g_{j,k}^{(l)}$. In arriving at this Hamiltonian, a rotating wave approximation (RWA) has been made and we have assumed that the transmon charge operator that gives rise to the interaction Hamiltonian only allows appreciable coupling between nearest-neighbor energy levels \cite{koch2007charge}.

The next step in a first-principles derivation of the Maxwell-Schr\"{o}dinger framework is to make a displacement transformation on the EM part of the system \cite{vsindelka2010derivation}. Such displacement transforms are common in quantum optics to transform a vacuum state to a coherent state \cite{walls2008quantum}, and in our context here is necessary since a semiclassical Maxwell-Schr\"{o}dinger method will naturally need to assume that the EM part of the system is in a coherent state. In particular, we perform a time-dependent displacement transform given by the operator
\begin{align}
     \hat{D}( \{\alpha_k (t)\} )=e^{\sum_k \alpha_k(t) \hat{a}_k^\dag-\alpha_k^*(t) \hat{a}_k}
     \label{eq:displacement-operator}
\end{align}
to account for the changing location of the coherent state in phase space due to the changing $\alpha_k$'s. These changes could be due to the impact of a drive by a voltage source, but can also be thought of as occurring due to the back-action that the transmons will have on the EM subspace. Applying this displacement transformation, our Hamiltonian in (\ref{EQ:Two-Qubit-Ham}) becomes
\begin{multline}
    \hat{H}_\mathrm{2D}= D(\{\alpha_k (t)\})^\dag \hat{H}_{2} D(\{\alpha_k (t)\})  \\ -i\hbar  D(\{\alpha_k (t)\})^\dag  \partial_t D(\{\alpha_k (t)\}),
\end{multline}
which evaluates to
\begin{multline}
    \label{EQ:Two-Qubit-Ham-Disp}
     \hat{H}_\mathrm{2D}= \hbar\sum_{j}\sum_{l=1}^{2}q_{j}^{(l)}\ket{j}^{(l)}\bra{j}^{(l)}\\+\hbar\sum_{k} \omega_k\biggl[\hat{a}_{k}^\dag \hat{a}_{k}+\hat{a}_{k}^\dag\alpha_{k}+\hat{a}_{k}\alpha^{*}_{k}+|\alpha_{k}|^2\biggr]\\+
        \hbar\sum_{j,k} \left[g_{j,k}^{(l)}\ket{j}^{(l)}\bra{j+1}^{(l)}(\hat{a}_k^\dag+\alpha^*_k)+\textnormal{H.c}\right]\\
        -i\hbar\left[ \sum_{k} \dot{\alpha}_k(\hat{a}^\dag_k+\alpha^*_k)-\dot{\alpha}^*_k(\hat{a}_k+\alpha_k)\right],
\end{multline}
where $\dot{\alpha}_k \equiv \partial_t \alpha_k$ and we have suppressed the dependence on $t$ to simplify the notation.


This transformation has the benefit of separating the effects of the ``classical drive'' (the $\alpha_k$ terms) and the quantum fluctuations (the $\hat{a}_k$ and $\hat{a}^\dag_k$ terms). However, this form of the Hamiltonian is not directly useful in identifying the multi-qubit effects as these are still expressed indirectly through the interactions of the individual transmons with the common EM system. Typically, cQED devices operate in the dispersive regime where the qubit transition frequencies are significantly detuned from any EM resonator modes \cite{blais2021circuit}. When this is the case, we can perform a Schrieffer-Wolff transformation to eliminate the explicit qubit-EM interaction terms and arrive at an effective Hamiltonian that directly displays the qubit-qubit interaction terms. The details of this transformation are given in Appendix \ref{sec:appendix-SW}, with the final result denoted as $\hat{H}_\mathrm{2DS}$ which is given in (\ref{Eq:Two-SW}). 

For our purposes here, we further assume that the EM subspace remains only in a coherent state so that the quantum fluctuations that need to be captured in our equations are just due to vacuum fluctuations. We can focus only on these relevant terms by projecting (\ref{Eq:Two-SW}) into the zero-excitation subspace of the EM system. Denoting this zero-excitation subspace with the state $\ket{0}_\mathrm{R}$, our projected Hamiltonian becomes
\begin{widetext}
    \begin{multline}
    \label{EQ:Two-Qubit-Ham-Disp-SW}
    \bra{0}_\mathrm{R}\hat{H}_\mathrm{2DS}\ket{0}_\mathrm{R}=\hbar\sum_{j}\sum_{l=1}^{2}q_{j}^{(l)}\ket{j}^{(l)}\bra{j}^{(l)}
    +\sum_{k} \hbar \omega_k|\alpha_k|^2-i\hbar \sum_{k} \left[\dot{\alpha}_k\alpha^*_k-\dot{\alpha}^*_k\alpha_k\right] 
    +\hbar \sum_{j,k}\sum_{l=1}^2\chi_{j,k}^{(l)}\ket{j+1}^{(l)}\bra{j+1}^{(l)} \\
    +\hbar\sum_{j,k}\sum_{l=1}^2\biggl[g_{j,k}^{(l)}\ket{j}^{(l)}\bra{j+1}^{(l)}(\alpha^*_k)     +\textnormal{H.c.}\biggr] +
    \hbar\sum_{j,k}\sum_{l=1}^2\biggl[\big(B_{j,k}^{(l)} \big)^*\ket{j+1}^{(l)}\bra{j}^{(l)}( \omega_k\alpha_k-i \dot{\alpha}_k)+\textnormal{H.c.}\biggr]\\
    +\hbar\sum_{i,j} J_{ij}\biggl[\ket{i}^{(1)}\bra{i+1}^{(1)}\ket{j+1}^{(2)}\bra{j}^{(2)} +\ket{i+1}^{(1)}\bra{i}^{(1)}\ket{j}^{(2)}\bra{j+1}^{(2)}\biggr],
\end{multline}
\end{widetext}
where 
\begin{align}
    \chi_{j,k}^{(l)}=\frac{\lvert {g}_{j,k}^{(l)}\rvert ^2}{q_{j,j+1}^{(l)}-\omega_k},
\end{align}
\begin{align}
    B_{j,k}^{(l)}=\frac{{g}_{j,k}^{(l)}}{q_{j,j+1}^{(l)}-\omega_k},
\end{align}
\begin{align}
    \label{EQ:J-Coupling}
    J_{ij}= \sum_k \frac{1}{2} \left[\frac{{g}_{i,k}^{*(1)}{g}_{j,k}^{(2)}}{(q_{i,i+1}^{(1)}-\omega_k)}+\frac{{g}_{i,k}^{(1)}{g}_{j,k}^{*(2)}}{(q_{j,j+1}^{(2)}-\omega_k)}\right],
\end{align}
and $q_{j,j+1}^{(l)} = q_{j+1}^{(l)} - q_{j}^{(l)}$. Inspecting (\ref{EQ:Two-Qubit-Ham-Disp-SW}), we can identify typical terms due to the Lamb shift on the third line, the effects of the applied drive or qubit back-action on the fourth and fifth lines, and the multi-qubit exchange coupling terms due to $J_{ij}$. 


Critically, as shown in Appendix \ref{app:single-qubit}, when we repeat this analysis for only a single transmon coupled to an EM system, we find that all the same terms appear with the exception of the multi-qubit exchange coupling terms. Since the Maxwell-Schr\"{o}dinger method has been shown to quantitatively capture single-qubit effects in \cite{roth2024maxwell,elkin2024epeps} and Appendix \ref{sec:Open-Quant}, we may further conclude that all terms that depend on the $\alpha_k$'s in (\ref{EQ:Two-Qubit-Ham-Disp-SW}) are naturally captured by the Maxwell-Schr\"{o}dinger method and require no modifications to that numerical method to correctly model these kinds of interactions for the multi-qubit case. Hence, we only need to augment the Maxwell-Schr\"{o}dinger method to explicitly account for the multi-qubit exchange coupling terms. These modifications will be explicitly given in Section \ref{subsec:multi-qubit-Maxwell-Schrodinger}.

However, doing this necessitates an accurate way to compute $J_{ij}$, which is complicated due to the summation over all the modes of the EM system. Evaluating this expression explicitly is problematic due to the computed value of $J_{ij}$ often changing appreciably whether the summation is truncated at an even or odd number of modes \cite{filipp2011multimode}, and more generally, because even computing the properties of these modes for a 3D simulation can be numerically challenging \cite{elkin2025opportunities}. However, these issues can be avoided by following the process of our prior work \cite{khan2024field} that shows how (\ref{EQ:J-Coupling}) can be directly related to the impedance parameters $Z_{lm}$ of the EM system, which can be modeled much more efficiently than extracting EM modes \cite{elkin2025opportunities}. In particular, \cite{khan2024field} shows that
\begin{multline}
\label{EQ:J-Coupling-Paper}
            J_{ij}=2\frac{e^2}{\hbar}
        \biggl[{n}_{i+1,i}^{(1)}{n}_{j,j+1}^{(2)} q_{i,i+1}^{(1)}\Im{ Z_{12}(q_{i,i+1}^{(1)})}
        \\+{n}_{j+1,j}^{(2)} {n}_{i,i+1}^{(1)}q_{j,j+1}^{(2)}\Im{Z_{21}(q_{j,j+1}^{(2)})}\biggr],
\end{multline}
where we have used the shorthand notation
\begin{align}\label{Def n term}
    n_{i,j}^{(l)} = \bra{i}^{(l)}\hat{n}^{(l)}\ket{j}^{(l)},
\end{align}
to simplify the expressions.

\subsection{Multi-Qubit Exchange Coupling in the Maxwell-Schr\"{o}dinger Method}
\label{subsec:multi-qubit-Maxwell-Schrodinger}
Here, we explicitly give the multi-qubit Maxwell-Schr\"{o}dinger equations of motion based on the formulation from the previous section. To simplify the notation, we will consider explicitly the two-qubit case; however, the generalization to more qubits is straightforward. Since there will be multiple qubits present, it makes more sense to write the Schr\"{o}dinger equation part in terms of the eigenstates of the individual qubit Hamiltonians rather than the phase basis that was used in (\ref{eq:schro-wave4}). In this case, the Hamiltonian for the Schr\"{o}dinger equation is
\begin{multline}
   \hat{H} = \hbar\sum_{j}\sum_{l=1}^{2} \bar{q}_{j}^{(l)}\ket{j}^{(l)}\bra{j}^{(l)}  \\ 
    +\hbar\sum_{i,j} J_{ij}\biggl[\ket{i}^{(1)}\bra{i+1}^{(1)}\ket{j+1}^{(2)}\bra{j}^{(2)}\\+\ket{i+1}^{(1)}\bra{i}^{(1)}\ket{j}^{(2)}\bra{j+1}^{(2)}\biggr] \\ + \sum_m \sum_{l=1}^{2} 2e\beta^{(l)}_m \big(  \partial_t \phi(z^{(l)}_m,t)  \big)  \hat{n}^{(l)},
    \label{eq:MS-2-qubit-Schro}
\end{multline}
where $\bar{q}_{j}^{(l)}$ has absorbed the Lamb shift into the definition of the qubit eigenfrequencies, $\beta^{(l)}_m$ is the capacitive voltage divider from the $m$-th transmission line to the $l$-th qubit, and $z^{(l)}_m$ is the coupling point from the $m$-th transmission line to the $l$-th qubit. As suggested by the form of (\ref{eq:MS-2-qubit-Schro}), it is most natural to discretize this equation in the tensor product basis of states of the form $\ket{ij} = \ket{i}^{(1)}\otimes \ket{j}^{(2)}$.

Similarly, the basic equation governing the node flux on the $m$-th transmission line becomes
\begin{align}
	\partial_z^2 \phi - L_m C_m \partial_t^2\phi = - \sum_{l=1}^{2} \delta(z-z^{(l)}_m) L_m 2e\beta^{(l)}_m \partial_t \langle {n}^{(l)} (t)\rangle,
	\label{eq:phi-coupled-wave-2-qubit}
\end{align}
where $L_m$ and $C_m$ are the per-unit-length inductance and capacitance of the line and $\langle {n}^{(l)} (t)\rangle$ is the expectation value of $\hat{n}^{(l)}$. The inclusion of additional voltage sources, capacitive coupling between different transmission lines, and resistive terminations can all be handled in a straightforward manner following the details in \cite{roth2024maxwell}. Here, we will discretize (\ref{eq:phi-coupled-wave-2-qubit}) using the same FEM approach as detailed in \cite{roth2024maxwell}.

Finally, the same leap-frog time-marching strategy discussed in Section \ref{sec:maxwell-schrodinger-background} and in \cite{roth2024maxwell} can be used to solve (\ref{eq:MS-2-qubit-Schro}) and (\ref{eq:phi-coupled-wave-2-qubit}) together. As we will show in Section \ref{sec:results}, this resulting Maxwell-Schr\"{o}dinger method will be able to correctly exhibit the expected multi-qubit dynamics. It will further aid in modeling the classical crosstalk effects that have been noted in prior cross-resonance gate experiments \cite{sheldon2016procedure}.

%% file: Numerical_Results.tex
\section{Numerical Results}
\label{sec:results}
\begin{figure}[t]
    \centering
    \includegraphics[width=\linewidth]{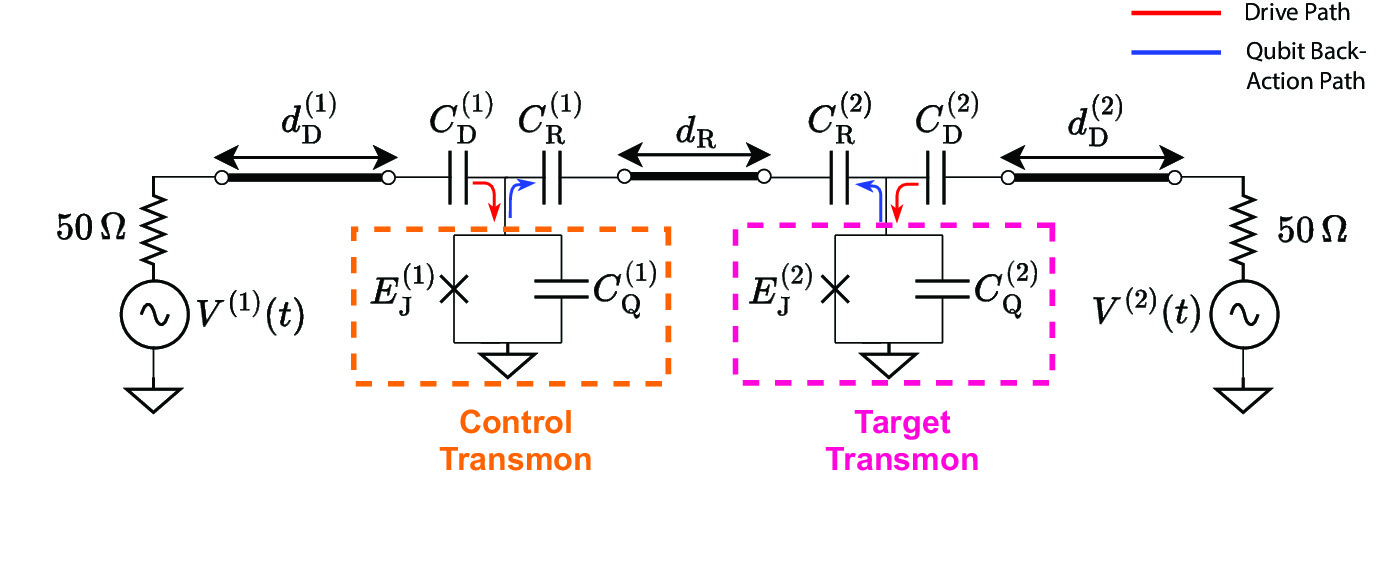}
    \caption{Device schematic with two transmons coupled capacitively to a common transmission line resonator of length $d_\mathrm{R}$. Each transmon is also connected to an independent drive source through additional transmission lines. This circuit enables a cross-resonance gate between the transmons, where for our simulations, Transmon 1 serves as the ``control transmon'' and Transmon 2 is the ``target transmon''.}
    \label{fig:two-qubit-schematic}
\end{figure}

To test our multi-qubit Maxwell-Schr\"{o}dinger model, we focus on the circuit shown in Fig. \ref{fig:two-qubit-schematic} that is designed to enable a cross-resonance gate between the two transmon qubits. This gate is widely used in academic \cite{rigetti2010fully,li2024experimental} and industry \cite{sheldon2016procedure,kandala2021demonstration} settings to achieve a microwave-only two-qubit interaction. In the device of Fig. \ref{fig:two-qubit-schematic}, the $j$-th transmon is coupled to a common transmission line resonator through capacitances $C_\mathrm{R}^{(j)}$. The transmons are also coupled to independent drive lines through capacitances $C_\mathrm{D}^{(j)}$. We focus our initial results on exploring the basic cross-resonance effect and the source of classical crosstalk in these devices due to the back-action of the transmons on the transmission line resonator. Following this, we consider additional simulations to demonstrate how this back-action effect can change substantially due to changes in the device parameters. 

For all simulations, we keep the coupling capacitances to the drive lines fixed at $C_{D}^{(1)}=C_{D}^{(2)}=0.1\;\mathrm{fF}$. The total transmon capacitances are also kept fixed at $C_{\Sigma}^{(1)}=67.95\;\mathrm{fF}$ and $C_{\Sigma}^{(2)}=67.45\;\mathrm{fF}$ so that the charging energy of our transmons remains the same for all our simulations. The values of $C_\mathrm{R}^{(j)}$, $C_\mathrm{Q}^{(j)}$, and $E_\mathrm{J}^{(j)}$ will be changed to demonstrate different simulations of interest. All simulations will also be conducted in the dispersive regime by keeping the first resonant frequency of the transmission line resonator set at $6.31\;\mathrm{GHz}$. This is accomplished by making $d_\mathrm{R}=5.66\;\mathrm{mm}$ with a per-unit-length capacitance of $C=280\;\mathrm{pF/m}$, and per-unit-length inductance of $L=0.7\;\mathrm{\mu H/m}$. These same $C$ and $L$ are used for all the other transmission lines as well.

While discretizing the circuit of Fig. \ref{fig:two-qubit-schematic} is easy using the Maxwell-Schr\"{o}dinger framework, with extensive details on the this process available in \cite{roth2024maxwell}, exactly replicating the impact of the various circuit connections in the typical fully-quantum models that are considered in this section can be laborious. Here, we make two simplifications within the Maxwell-Schr\"{o}dinger model to make the quantitative comparison with the fully-quantum model easier. First, we remove any direct capacitive coupling between the resonator and drive transmission lines, meaning that the applied voltage drives only affect the transmons and that there is no residual leakage directly from the drive lines to the resonator. Second, we only allow the back-action of the transmons to affect the resonator so that we can isolate the impact of this effect in comparing between the Maxwell-Schr\"{o}dinger and fully-quantum system models. These two simplifications are shown by the red and blue arrows in Fig. \ref{fig:two-qubit-schematic}.

Now, the system Hamiltonian appropriate for the circuit in Fig. \ref{fig:two-qubit-schematic} is similar to that of (\ref{eq:MS-2-qubit-Schro}), but for our purposes here will be
\begin{multline}
    \label{Eq:H_J}
    \hat{H}_J=\hbar\sum_{j}\sum_{l=1}^{2} \bar{q}_{j}^{(l)}\ket{j}^{(l)}\bra{j}^{(l)}  \\ 
    +\hbar\sum_{i,j} J_{ij}\biggl[\ket{i}^{(1)}\bra{i+1}^{(1)}\ket{j+1}^{(2)}\bra{j}^{(2)}\\+\ket{i+1}^{(1)}\bra{i}^{(1)}\ket{j}^{(2)}\bra{j+1}^{(2)}\biggr]. 
\end{multline}
To perform a cross-resonance pulse, the control transmon needs to be driven with a pulse that is resonant with the transition frequency of the target transmon \cite{krantz2019quantum,sheldon2016procedure}. Depending on the state of the control transmon, the frequency of the Rabi oscillations that correspondingly occur on the target transmon will change. This external drive is modeled here as
\begin{align}
    \label{Eq:H_d1}
    \hat{H}_{d1}=\hbar 2e\beta^{(1)} V^{(1)}(t,\phi)\hat{n}^{(1)},
\end{align}
where $V^{(1)}(t,\phi)$ is typically a modulated flat-top Gaussian pulse \cite{sheldon2016procedure} of the form
\begin{align}
    \label{Eq:flatop-Gaussian}
    V^{(1)}(t,\phi)=V_\mathrm{mag}\cos(2\pi f_qt+\phi)f(t).
\end{align}
Here, $V_\mathrm{mag}$ is the magnitude of the pulse, $\phi$ is a phase offset, and $f_q$ is the first transition frequency of the target transmon. Further, the pulse envelope is of the form
\[
f(t) =
\begin{cases}
\exp\biggl(-\frac{1}{2}\frac{(t-t_{0}-\zeta)^2}{\sigma^2}\biggr), & t < \zeta+t_0 \\
1,   &\zeta+t_0\leq t \leq \tau-\zeta+t_0\\
\exp\biggl(-\frac{1}{2}\frac{(t-t_{0}-\tau+\zeta)^2}{\sigma^2}\biggr), & \tau-\zeta+t_0 <t,
\end{cases}
\]
where $t_0$ is the time-offset, $\tau$ is the duration of the pulse, $\zeta$ is the rise-fall time, and $\sigma$ is the standard deviation of the Gaussian edges of the envelope. 

\begin{figure}[t]
    \centering
    \includegraphics[width=0.9\linewidth]{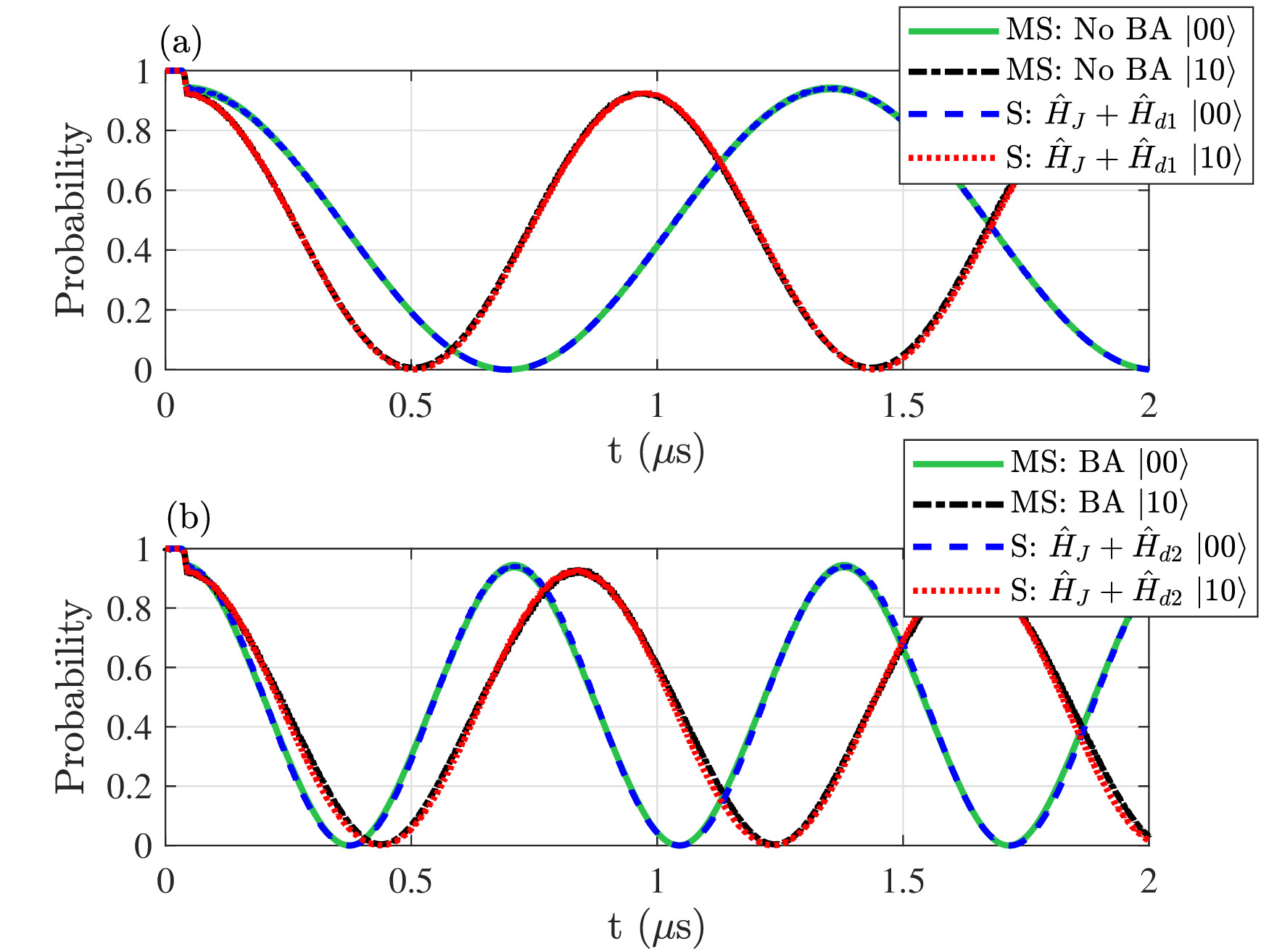}
    \caption{Simulation results of the basic cross-resonance effect. On each plot, results from two different simulations are shown, corresponding to initial states of $\ket{00}$ or $\ket{10}$. Depending on the state of the control transmon, the Rabi oscillation frequency for the target transmon changes. (a) Results are plotted for our Maxwell-Schr\"{o}dinger framework with back-action artificially turned off (MS: No BA) and for the standard time evolution using QuTiP of the total Hamiltonian $\hat{H}_J + \hat{H}_{d1}$. (b) The same simulations are repeated but with the Maxwell-Schr\"{o}dinger framework with back-action turned on (MS: BA) and for the standard time evolution using QuTiP of the total Hamiltonian $\hat{H}_J + \hat{H}_{d2}$ that includes a classical crosstalk term. As noted in previous experiments \cite{sheldon2016procedure,magesan2020effective}, the presence of a classical crosstalk term can significantly modify the cross-resonance effect.}
    \label{sec3a:Image:Uni_Rabi}
\end{figure}

For our first set of results, we set the parameters to have $V_\mathrm{mag}=140\;\mathrm{\mu V}$, $\sigma=5\;\mathrm{ns}$, $\zeta=15 \;\mathrm{ns}$, $\phi=0$, $\tau=2\;\mathrm{\mu s}$, and $t_0=30\;\mathrm{ns}$. The coupling capacitance to the resonator are both set at $C_\mathrm{R}^{(1)}=C_\mathrm{R}^{(2)}=4\;\mathrm{fF}$, and the Josephson energies are varied to get $\bar{q}_{0,1}^{(1)}=4.91\;\mathrm{GHz}$, and $\bar{q}_{0,1}^{(2)}=5.11\;\mathrm{GHz}$. The multi-qubit coupling rate $J_{ij}$ is evaluated using the method from \cite{khan2024field}. Instead of reporting the entire coupling matrix, we will report $J_{00}$, which is commonly referred to as $J$ in the literature, and is evaluated to be $1.63\;\mathrm{MHz}$ in this case. In Fig. \ref{sec3a:Image:Uni_Rabi}(a), we plot the results from the typical time evolution of the combined Hamiltonian from $\hat{H}_J + \hat{H}_{d1}$ (performed in QuTiP \cite{johansson2012qutip}) and our Maxwell-Schr\"{o}dinger method with qubit back-action artificially turned off, which is accomplished by removing the terms on the right-hand side of (\ref{eq:phi-coupled-wave-2-qubit}). The simulations are performed with the control transmon initialized in the $\ket{0}$ or $\ket{1}$ state, and we see the difference in the Rabi oscillation frequency of the target transmon characteristic of the cross-resonance effect \cite{krantz2019quantum}. We further see excellent agreement between both methods, showing that the voltage sources are calibrated appropriately between the two simulation methods.

As shown in Appendix \ref{sec:Open-Quant} for single-qubit gates, the qubit back-action on the transmission line system is important for accurately evaluating the quantum dynamics, particularly for pulses with longer durations. Compared to the single-qubit case, the role of this qubit back-action is even more essential when considering the two-qubit case as it can be a source of classical crosstalk. Such classical crosstalk has been predicted to be responsible for significant effects that cannot be explained through quantum analysis alone \cite{sheldon2016procedure,magesan2020effective}, which is why modeling it robustly is essential towards improving cQED designs. In \cite{magesan2020effective}, a simple strategy was used to consider this classical crosstalk effect by altering (\ref{Eq:H_d1}) to include an additional scaled drive directly on the target transmon with a different phase factor. The corresponding drive Hamiltonian then becomes
\begin{multline}
    \label{Eq:H_d2}
    \hat{H}_{d2}=\hbar2e\beta^{(1)}V^{(1)}(t,\phi)\hat{n}^{(1)} \\ +\hbar A2e\beta^{(1)}V^{(1)}(t,\phi_2)\hat{n}^{(2)},
\end{multline}
where $A$ scales the amplitude and $\phi_2$ allows for an additional phase shift. Such additional drives are then typically tuned to match experimental results to gain a sense of the amount of classical crosstalk present in the system. By including back-action into our Maxwell-Schr\"{o}dinger method, we can incorporate this classical crosstalk naturally into our simulation and see its impact on the qubit dynamics. We plot these results in Fig. \ref{sec3a:Image:Uni_Rabi}(b) along with the typical time evolution of the combined Hamiltonians $\hat{H}_J + \hat{H}_{d2}$. We use the fitting parameters $A=0.007$ and $\phi_2=0$ when the control transmon is in the $\ket{0}$ state, and $A=0.0018$ and $\phi_2=\pi$ when the control transmon is in the $\ket{1}$ state. We see good agreement in the overall Rabi frequencies, although the dynamics do vary somewhat between the methods over the full pulse.

\begin{figure}[t]
    \centering
    \includegraphics[width=0.9\linewidth]{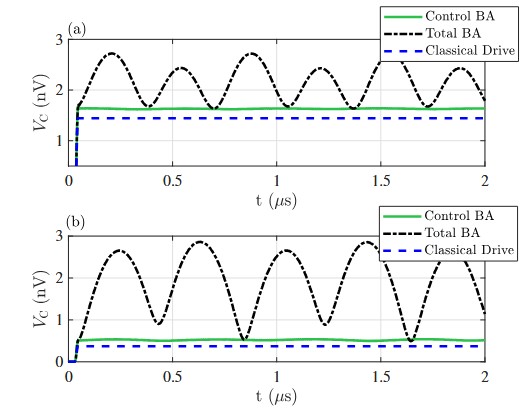}
    \caption{The envelope of the voltage $V_\mathrm{C}(t)$ that is responsible for the classical crosstalk in Fig. \ref{sec3a:Image:Uni_Rabi}. This voltage is naturally evaluated in the Maxwell-Schr\"{o}dinger framework as the back-action of the control and target transmons into the transmission line resonator. We monitor this voltage for simulations when only the back-action of the control transmon is considered (Control BA) and when both the control and target transmons back-action are considered (Total BA). These are compared against the envelope of the external classical drive that was required to make the results of the drive Hamiltonian $\hat{H}_{d2}$ in our quantum simulation in Fig. \ref{sec3a:Image:Uni_Rabi}(b) match the Maxwell-Schr\"{o}dinger results. These envelopes are shown for initial states of (a) $\ket{00}$ and (b) $\ket{10}$.} 
    \label{sec3a:Image:Rabi_envelope}
\end{figure}

To investigate the back-action effect further, we monitor the voltage in the Maxwell-Schr\"{o}dinger simulations at the end of the resonator that is coupled to the target transmon for several scenarios and plot their envelopes in Fig. \ref{sec3a:Image:Rabi_envelope}(a). We first look at the case when the control transmon starts in the $\ket{0}$ state and where back-action from only the control transmon into the resonator is considered. We see that this closely resembles the envelope of the classical crosstalk drive assumed in (\ref{Eq:H_d2}) that is based on \cite{magesan2020effective}. However, we see that when we consider the total back-action from both transmons, the voltage envelope is clearly quite different. While the assumed simple form of crosstalk envelope in $\hat{H}_{d2}$ is able to reasonably match the qubit dynamics in Fig. \ref{sec3a:Image:Uni_Rabi}(b), such predictions may not be sufficient for all applications given how different the voltage envelopes can be. We also repeat these tests in  Fig. \ref{sec3a:Image:Rabi_envelope}(b) for the control qubit in the $\ket{1}$ state, which clearly shows that the back-action effect is different. This further suggests that tuning additional classical drive models with these simple pulse envelopes could become insufficient for more complicated devices and more advanced pulse designs.

We now move to illustrating how the qubit dynamics change due to the qubit back-action as we change several key circuit parameters. In all the following simulations, the drive duration is set at $\tau=340\;\mathrm{ns}$ and we simulate the dynamics for $400\;\mathrm{ns}$. The rise-fall times and Gaussian standard deviations remain the same as the previous simulations, as does the $V_\mathrm{mag}$. 

\begin{figure}[t]
    \centering
    \includegraphics[width=0.9\linewidth]{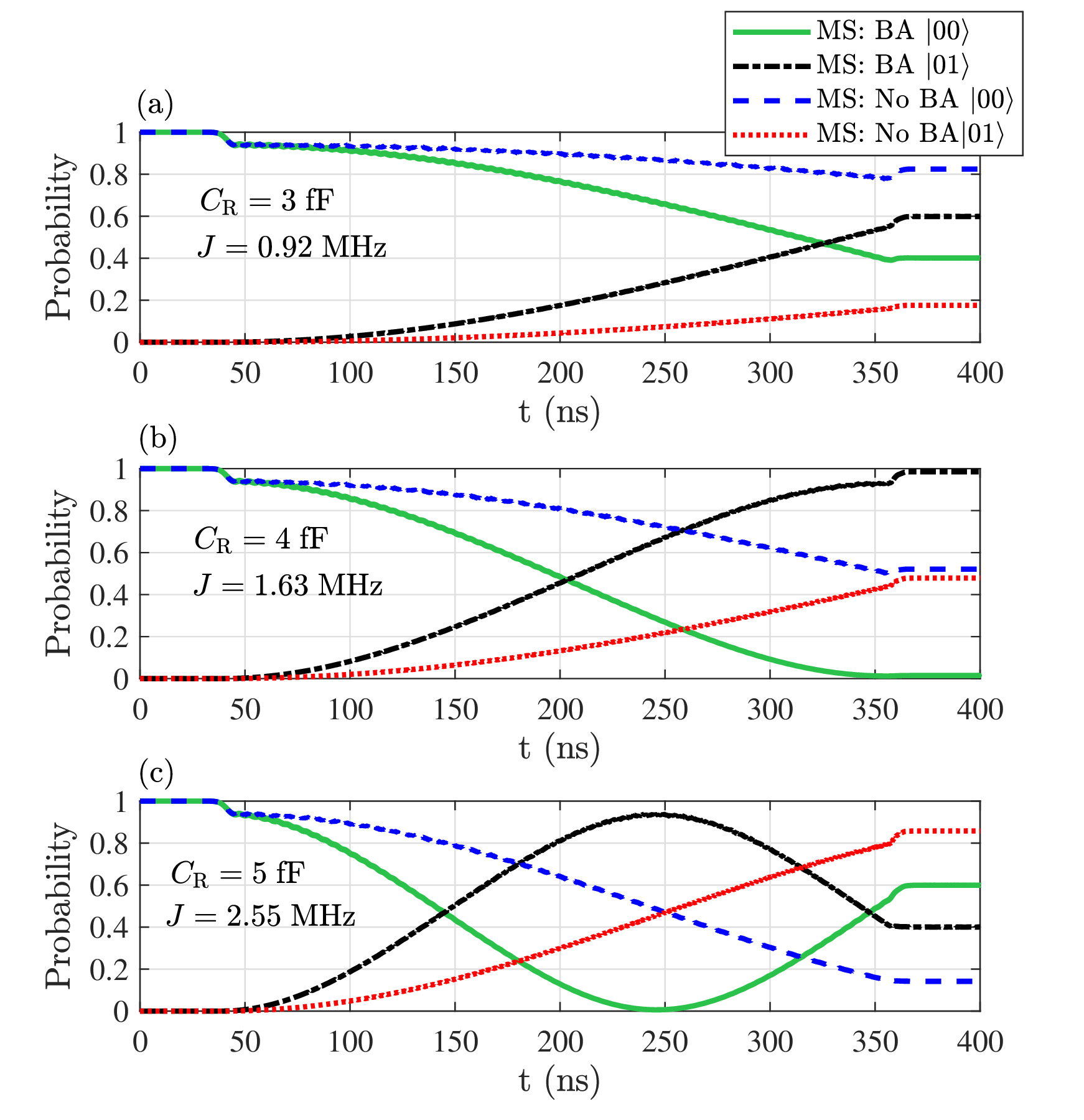}
    \caption{Maxwell-Schr\"{o}dinger simulations with back-action (MS: BA) and without (MS: No BA) for the circuit of Fig. \ref{fig:two-qubit-schematic} while varying the coupling capacitances to the resonator. Each simulation starts in the $\ket{00}$ state and we apply a $340\;\mathrm{ns}$ duration pulse. The device is symmetric so that $C_\mathrm{R}^{(1)}=C_\mathrm{R}^{(2)}=C_\mathrm{R}$ and $C_\mathrm{D}^{(1)}=C_\mathrm{D}^{(2)}=0.1\;\mathrm{fF}$. Qubit frequencies are fixed at $\bar{q}_{0,1}^{(1)}/2\pi=4.91\;\mathrm{GHz}$ and $\bar{q}^{(2)}_{0,1}/2\pi=5.11\;\mathrm{GHz}$. The resonator coupling capacitances are (a) $C_\mathrm{R}=3\;\mathrm{fF}$, (b) $C_\mathrm{R}=4\;\mathrm{fF}$, and (c) $C_\mathrm{R}=5\;\mathrm{fF}$.}
    \label{sec3a:Image:Diff_C}
\end{figure}

The first set of results explores the impact of changing the coupling capacitance to the resonator, while keeping the circuit symmetric $C_\mathrm{R}^{(1)}=C_\mathrm{R}^{(2)}=C_\mathrm{R}$. In Fig. \ref{sec3a:Image:Diff_C}, we show the dynamics for systems with $C_\mathrm{R}=3\;\mathrm{fF},$ $4\;\mathrm{fF}$, and $5\;\mathrm{fF}$ with the qubits initialized to the $\ket{00}$ state while keeping all the other circuit settings the same as the previous simulations. These parameters are chosen to keep the qubit-qubit coupling rate $J$ to comparable levels as reported in the literature \cite{solgun2019simple}. We can expect that increased capacitance to the resonator will increase the classical crosstalk, which would increase the deviation between the dynamics predicted between the Maxwell-Schr\"{o}dinger methods with and without back-action. This is exactly the trend shown in Fig. \ref{sec3a:Image:Diff_C}.

\begin{figure}[t]
    \centering
    \includegraphics[width=0.9\linewidth]{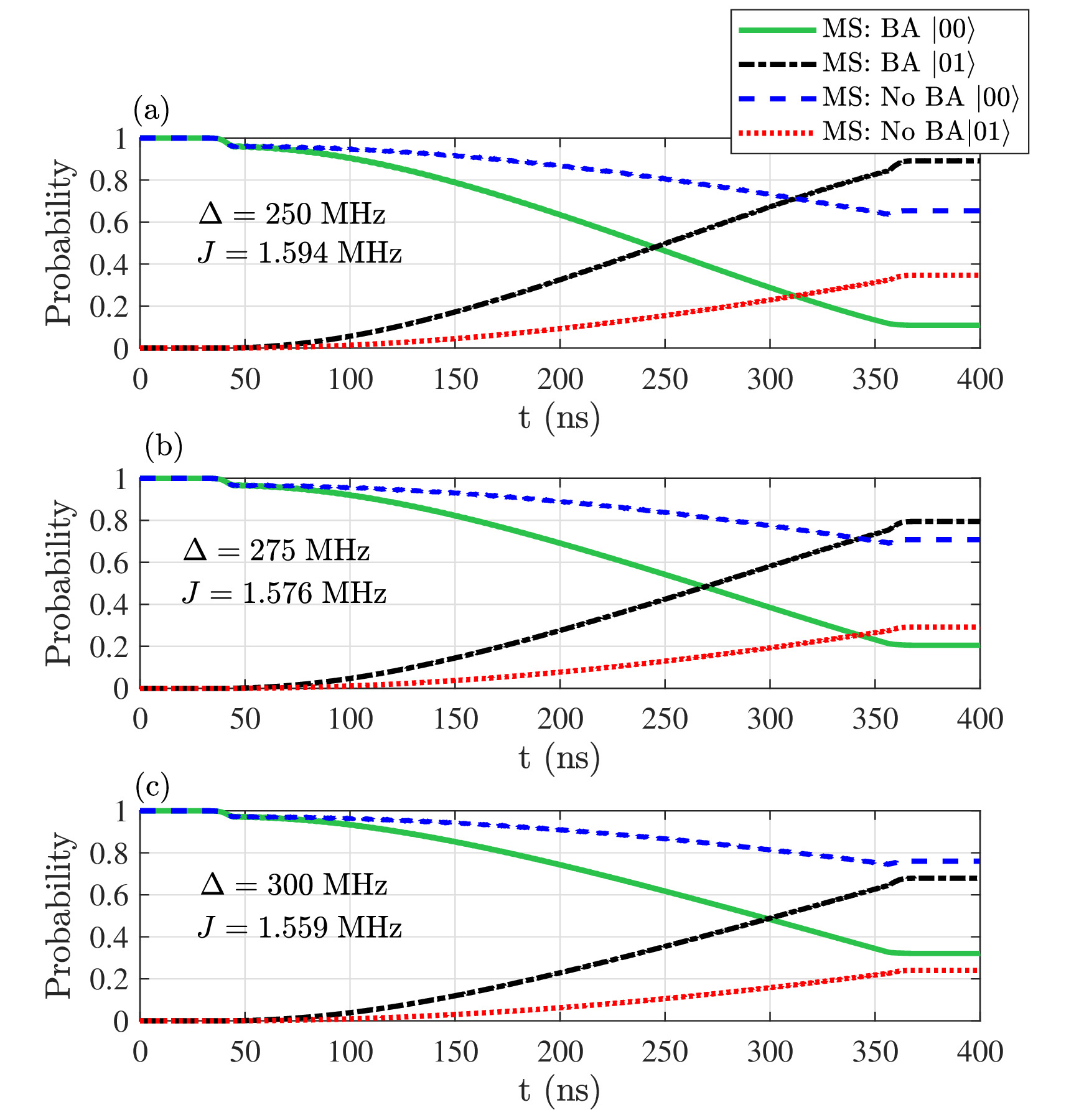}
    \caption{Maxwell-Schr\"{o}dinger simulations with back-action (MS: BA) and without (MS: No BA) for the circuit of Fig. \ref{fig:two-qubit-schematic} while varying the qubit-qubit detuning $\Delta=(\bar{q}_{0,1}^{(2)}-\bar{q}_{0,1}^{(1)})/2\pi$. Each simulation starts in the $\ket{00}$ state and we apply a $340\;\mathrm{ns}$ duration pulse. The device is symmetric with $C_\mathrm{R}^{(1)}=C_\mathrm{R}^{(2)}=4\;\mathrm{fF}$ and $C_\mathrm{D}^{(1)}=C_\mathrm{D}^{(2)}=0.1\;\mathrm{fF}$. The target transmon frequency is fixed at $\bar{q}^{(2)}_{0,1}/2\pi=5.11\;\mathrm{GHz}$, while we vary $E_\mathrm{J}^{(1)}$ to change the frequency of the control transmon for the different detunings. The different detunings are (a) $\Delta=250\;\mathrm{MHz}$, (b) $\Delta=275\;\mathrm{MHz}$, and (c) $\Delta=300\;\mathrm{MHz}$.}
    \label{sec3a:Image:Diff_Delta}
\end{figure}

Next, we look at the impact of changing the detuning between the control and target transmons, denoted by $\Delta=(\bar{q}_{0,1}^{(2)}-\bar{q}_{0,1}^{(1)})/2\pi$. The device is again kept symmetric with $C_\mathrm{R}^{(1)}=C_\mathrm{R}^{(2)}=4\;\mathrm{fF}$ and $C_\mathrm{D}^{(1)}=C_\mathrm{D}^{(2)}=0.1\;\mathrm{fF}$. The target transmon frequency is fixed at $\bar{q}^{(2)}_{0,1}/2\pi=5.11\;\mathrm{GHz}$, while we lower $E_\mathrm{J}^{(1)}$ to decrease the frequency of the control transmon to achieve different detunings. In Fig. \ref{sec3a:Image:Diff_Delta}, we plot the dynamics for the Maxwell-Schr\"{o}dinger methods with and without back-action for $\Delta=250\;\mathrm{MHz}$, $275\;\mathrm{MHz}$, and $300\;\mathrm{MHz}$. These changes slightly modify the multi-qubit coupling rate $J$ and have a moderate impact on the resulting classical crosstalk. This makes sense because the increase in qubit detunings also implies an increase in the detuning between the control transmon and the drive applied to it. Since this transmon is a weakly anharmonic oscillator, we can reasonably predict it to have a reduced back-action when the drive is further away from its own natural resonance frequency.

\begin{figure}[t]
    \centering
    \includegraphics[width=0.9\linewidth]{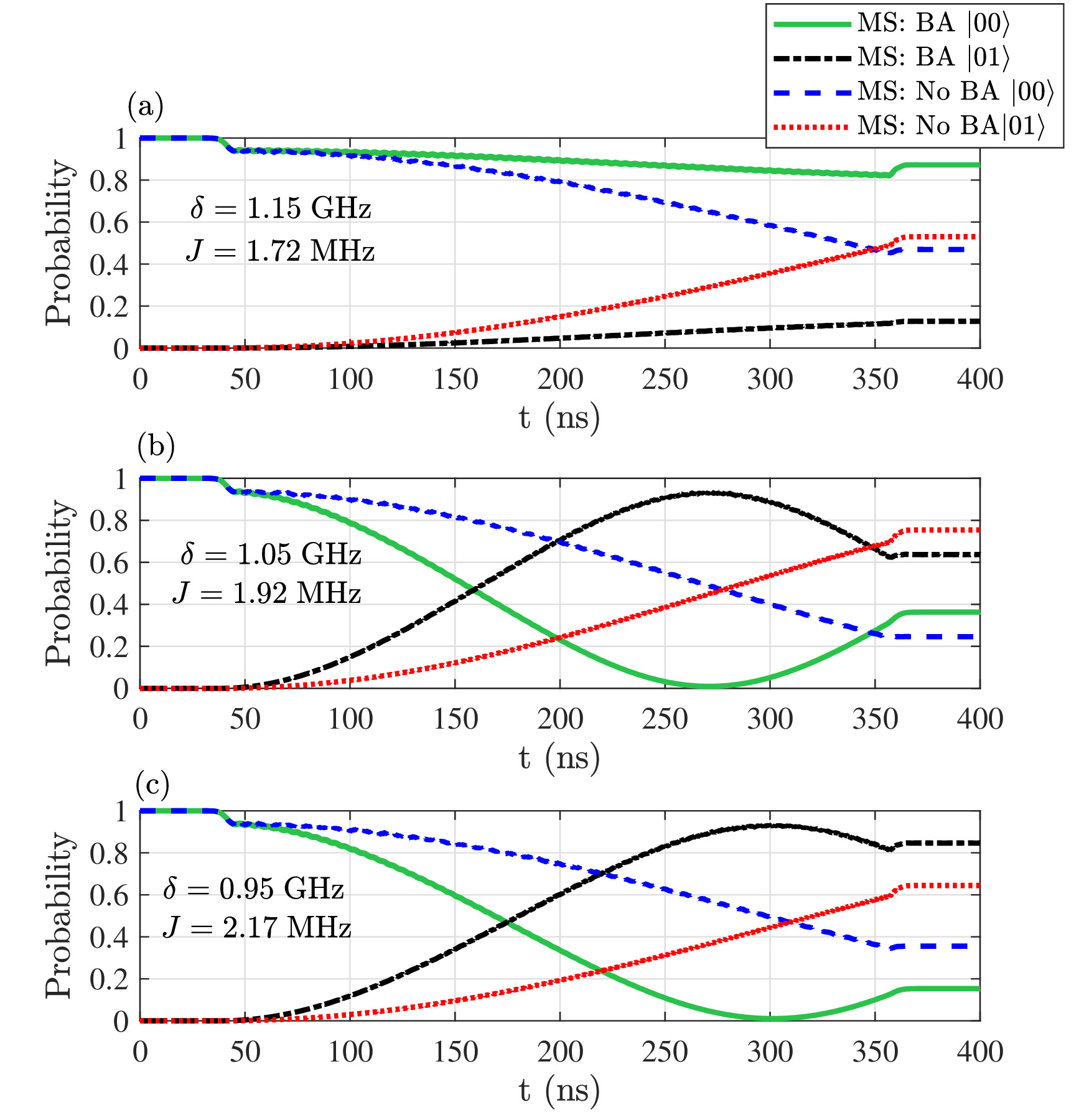}
    \caption{Maxwell-Schr\"{o}dinger simulations with back-action (MS: BA) and without (MS: No BA) for the circuit of Fig. \ref{fig:two-qubit-schematic} while varying the detuning for the target transmon with the resonator $\delta=(\omega_1-\bar{q}_{0,1}^{(2)})/2\pi$, while keeping the detuning between qubits constant at $\Delta=200\;\mathrm{MHz}$. The device is symmetric with $C_\mathrm{R}^{(1)}=C_\mathrm{R}^{(2)}=4\;\mathrm{fF}$ and $C_\mathrm{D}^{(1)}=C_\mathrm{D}^{(2)}=0.1\;\mathrm{fF}$. The different transmon-resonator detunings are (a) $\delta=1.15\;\mathrm{GHz}$, (b) $\delta=1.05\;\mathrm{GHz}$, and (c) $\delta=0.95\;\mathrm{GHz}$.}
    \label{sec3a:Image:Diff_Res}
\end{figure}

Lastly, we look at the impact of increasing both qubit frequencies closer to the first resonance of the transmission line resonator while still operating in the dispersive regime. We slowly decrease the resonator detuning $\delta=(\omega_1-\bar{q}_{0,1}^{(2)})/2\pi$, while keeping the detuning between transmons constant at $\Delta=200\;\mathrm{MHz}$. Again, the device is symmetric with $C_\mathrm{R}^{(1)}=C_\mathrm{R}^{(2)}=4\;\mathrm{fF}$ and $C_\mathrm{D}^{(1)}=C_\mathrm{D}^{(2)}=0.1\;\mathrm{fF}$. In Fig. \ref{sec3a:Image:Diff_Res}, we plot the dynamics for the Maxwell-Schr\"{o}dinger methods with and without back-action for $\delta=1.15\;\mathrm{GHz}$ $1.05\;\mathrm{GHz}$, and $0.95\;\mathrm{GHz}$. We see that sometimes the qubit back-action can increase or decrease the target transmon Rabi oscillation frequency, demonstrating the need to carefully account for it in designing cQED devices. 

%% file: Conclusion.tex
\section{Conclusion}
\label{sec:conclusion}
In this work, we have demonstrated how multi-qubit exchange coupling effects can be rigorously incorporated into Maxwell-Schr\"{o}dinger numerical methods. To support this, we built on a first-principles derivation of Maxwell-Schr\"{o}dinger methods for the specific case of two transmon qubits coupled to each other in the dispersive regime of cQED. We further provided a new interpretation of Maxwell-Schr\"{o}dinger methods as an efficient simulation strategy for capturing the class of non-Markovian open quantum system dynamics when the Born approximation holds and the EM circuitry is in a coherent state. From this perspective, we showed that our Maxwell-Schr\"{o}dinger methods can serve as a predictive simulation strategy to model the presence of strong classical crosstalk that has been found in experiments on cross-resonance gates, but for which previous quantum theory and device analysis could not explain their origin. Our Maxwell-Schr\"{o}dinger simulations demonstrate that the back-action of the qubits on the EM circuitry can provide this classical crosstalk, and that this effect can substantially alter the qubit dynamics. Our methods can be used in the future to better optimize the device and control pulse design to mitigate the influence of this deleterious classical crosstalk, helping improve the fidelity of multi-qubit gates in superconducting qubit processors.

To support this, Maxwell-Schr\"{o}dinger methods can be embedded into quantum optimal control methods to improve the realism of the underlying simulations utilized in the optimization methods. This is especially important for gradient- and machine learning-based optimization methods, as the success of translating these optimized pulses to hardware can greatly depend on the fidelity of the underlying simulation method \cite{jones2021approximations}. Further, our Maxwell-Schr\"{o}dinger methods can also be improved by leveraging a 3D full-wave solver for the Maxwell part of the simulations. This will make the methods more predictive to aid in design sign-off prior to fabrication. However, given the complex multiscale features of typical cQED devices \cite{elkin2025opportunities}, a robust 3D full-wave solver such as those investigated in \cite{elkin2025improvements} is likely to be needed. Additional improvements to our Maxwell-Schr\"{o}dinger framework include incorporating other open quantum system effects and leveraging tensor network compression so that larger numbers of qubits can be simulated efficiently.

%% file: Appendix_SQ.tex
\section{Quantum Formulation of Single-Qubit Maxwell-Schr\"{o}dinger Method}
\label{app:single-qubit}
Here, we follow the same process established in Section \ref{subsec:quantum-formulation} but apply it to a single-qubit model to show that many of the terms that appear in (\ref{EQ:Two-Qubit-Ham-Disp-SW}) are single-qubit effects that are already captured by a Maxwell-Schr\"{o}dinger method. We start with a single-qubit form of the Hamiltonian presented in (\ref{EQ:Two-Qubit-Ham}), which is
\begin{multline}
    \label{Sec2B:Eq:QH_full}
\hat{H}_{1}=\hbar\sum_{j} q_j\ket{j}\bra{j}+\hbar\sum_{k} \omega_k\hat{a}_k^\dag \hat{a}_k \\ +\hbar\sum_{j,k} \left[g_{j,k}\ket{j}\bra{j+1}\hat{a}_k^\dag+
\textnormal{H.c}\right].
\end{multline}

We now perform a time-dependent displacement transform using the operator from (\ref{eq:displacement-operator}) to get
\begin{multline}
    \hat{H}_\mathrm{1D}= D(\{\alpha_k (t)\})^\dag \hat{H}_{1} D(\{\alpha_k (t)\})  \\ -i\hbar  D(\{\alpha_k (t)\})^\dag  \partial_t D(\{\alpha_k (t)\}),
\end{multline}
which evaluates to
\begin{multline}
    \hat{H}_\mathrm{1D} = \hbar\sum_{j} q_j\ket{j}\bra{j} \\ +\hbar\sum_{k} \omega_k\biggl[\hat{a}_{k}^\dag \hat{a}_{k}+\hat{a}_{k}^\dag\alpha_{k}+\hat{a}_{k}\alpha^{*}_{k}+|\alpha|_{k}^2\biggr]\\+
        \hbar\sum_{j,k}\left[ g_{j,k}\ket{j}\bra{j+1}(\hat{a}_k^\dag+\alpha^*_k)+\textnormal{H.c}\right]\\
        -i\hbar \sum_{k} \left[\dot{\alpha}_k(\hat{a}^\dag_k+\alpha^*_k)-\dot{\alpha}^*_k(\hat{a}_k+\alpha_k)\right],
        \label{Eq:Single_Displaced}
\end{multline}
where we have suppressed the explicit time-dependence of the $\alpha_k$'s for brevity. Next, we can perform a Schrieffer-Wolff transformation to eliminate the direct qubit-EM interaction terms. The result of this is denoted as $\hat{H}_\mathrm{1DS}$ and is given in (\ref{Eq:Single-SW}) in Appendix \ref{sec:appendix-SW}, which once projected onto the zero-excitation subspace of the EM system gives
\begin{multline}
\label{Eq:Disp-SW-BA}
    \bra{0}_\mathrm{R}\hat{H}_\mathrm{1DS}\ket{0}_\mathrm{R}=\hbar\sum_{j} q_j\ket{j}\bra{j}
    +\sum_{k} \hbar \omega_k|\alpha_k|^2\\-i\hbar \sum_{k} \left[\dot{\alpha}_k\alpha^*_k-\dot{\alpha}^*_k\alpha_k\right]
    +\hbar\sum_{j,k}\chi_{j,k}\ket{j+1}\bra{j+1}\\
    +\hbar\sum_{j,k} \biggl[g_{j,k}\ket{j}\bra{j+1}(\alpha^*_k)     +\textnormal{H.c.}\biggr]\\
        +\hbar\sum_{j,k}\biggl[B_{j,k}^{*}\ket{j+1}\bra{j}( \omega_k\alpha_k-i \dot{\alpha}_k)+\textnormal{H.c.}\biggr].
\end{multline}

Comparing (\ref{Eq:Disp-SW-BA}) with (\ref{EQ:Two-Qubit-Ham-Disp-SW}), we see that the only type of terms in (\ref{EQ:Two-Qubit-Ham-Disp-SW}) that are not present in (\ref{Eq:Disp-SW-BA}) are those due to the multi-qubit exchange coupling. As mentioned in the main text, the results in \cite{roth2024maxwell,elkin2024epeps} and Appendix \ref{sec:Open-Quant} have shown that both single-qubit readout and control effects are captured well by the Maxwell-Schr\"{o}dinger method as described in (\ref{eq:schro-wave4}) and (\ref{eq:phi-coupled-wave}). Hence, we can conclude that all the $\alpha_k$-dependent terms in (\ref{Eq:Disp-SW-BA}) are accurately incorporated into the typical Maxwell-Schr\"{o}dinger interactions. This only leaves the multi-qubit exchange terms to then be added into the Hamiltonian used in the Schr\"{o}dinger equation to extend the Maxwell-Schr\"{o}dinger method to the multi-qubit case, as discussed in Section \ref{subsec:multi-qubit-Maxwell-Schrodinger}.

%% file: Appendix_SW.tex
\section{Schrieffer-Wolff Transformation}
\label{sec:appendix-SW}
The Schrieffer-Wolff transformation is used to find an effective Hamiltonian by removing the explicit interactions between weakly-coupled subspaces. In cQED literature, this transform has been widely used to find the effective impact of an EM resonator intentionally coupled to single qubit systems \cite{koch2007charge,zhu2013circuit} and to find effective qubit-qubit interactions in multi-qubit systems mediated by a common EM resonator \cite{filipp2011multimode,khan2024field,magesan2020effective}. The method is applied as a unitary transform to some Hamiltonian using the relation
\begin{align}\label{SW-Unitary}
\hat{H}_{\textrm{eff}}=e^{i\hat{S}}\hat{H}e^{-i\hat{S}},
\end{align}
where $\hat{S}$ is the generator. The details on finding and implementing this generator can be found in \cite{cohen1992atom}. Here, we focus on presenting the salient steps of the transform for the single- and two-qubit cases considered in Section \ref{sec:multi-qubit} and Appendix \ref{app:single-qubit}.

Beginning with the single qubit case, we wish to perform the Schrieffer-Wolff transformation on the displacement transformed Hamiltonian given in (\ref{Eq:Single_Displaced}). Using standard techniques, the generator can be given as
\begin{align}\label{Generator}
    \hat{S}=-i\Bigg[\sum_{j,k} B_{j,k}^{*}\ket{j+1}^{}\bra{j}^{}\hat{a}_k- \textnormal{H.c.}\Bigg]   
\end{align}
where
\begin{align}
    B_{j,k}=\frac{{g}_{j,k}}{q_{j,j+1}-\omega_k}.
\end{align}
Performing the transformation in (\ref{SW-Unitary}) up to the second-order in the coupling parameter $g_{j,k}$, we get the following effective Hamiltonian 
\begin{widetext}
    \begin{multline}\label{Eq:Single-SW}         
    \hat{H}_{\mathrm{1DS}}=\hbar\sum_jq_j\ket{j}\bra{j}+\hbar\sum_k \omega_k\biggl[\hat{a}_k^\dag \hat{a}_k+|\alpha_k|^2\biggr]
    +\hbar\sum_{j,k} \biggl[g_{j,k}\ket{j}\bra{j+1}(\alpha^*_k)+\textnormal{H.c.}\biggr] \\
         +\hbar\sum_k \biggl[\hat{a}^\dag_k( \omega_k\alpha_k-i \dot{\alpha}_k)+\hat{a}_k( \omega_k\alpha^*_k+i\dot{\alpha}^*_k)-i(\alpha^*_k-\alpha_k)\biggr]
         +\hbar\sum_{j,k}\left[\chi_{j,k}\ket{j+1}\bra{j+1}(\hat{a}_k^\dag \hat{a}_k+1)
         -\chi_{j,k}\ket{j}\bra{j}\hat{a}_{k}^\dag \hat{a}_k\right] \\
         +\hbar\sum_{j,k} \biggl[\chi_{j,k}\ket{j+1}\bra{j+1}\hat{a}_k\alpha^*_{k}
         +\textnormal{H.c.}\biggr]-\hbar\sum_{j,k} \biggl[\chi_{j,k}\ket{j}\bra{j}\hat{a}_k^\dag \alpha_{k}+\textnormal{H.c.}\biggr]+\hbar
        \sum_{j,k}\biggl[B_{j,k}^{*}\ket{j+1}\bra{j}( \omega_k\alpha_k-i \dot{\alpha}_k)+\textnormal{H.c.}\biggr],
     \end{multline}
\end{widetext}
where we have dropped terms of $O(B_{j,k}^2)$ and smaller and those corresponding to two-photon transitions, as is commonly done \cite{koch2007charge}. We also introduce the shorthand notation 
\begin{align}
    \chi_{j,k}^{}=\frac{\lvert {g}_{j,k}^{}\rvert ^2}{q_{j,j+1}^{}-\omega_k},
\end{align}
which correspond to terms due to the Lamb and ac-Stark shifts.

Now, for the two-qubit case, we can perform a similar Schrieffer-Wolff transformation to (\ref{EQ:Two-Qubit-Ham-Disp}) with the generator modified to
\begin{align}\label{Eq:Generator-2Q}
    \hat{S}=-i\Bigg[\sum_{j,k}\sum_{l=1}^2 \big(B_{j,k}^{(l)} \big)^* \ket{j+1}^{(l)}\bra{j}^{(l)}\hat{a}_k- \textnormal{H.c.}\Bigg].
\end{align}
Repeating the same steps as before, the effective Hamiltonian is found to be
\begin{widetext}
    \begin{multline}\label{Eq:Two-SW}         
    \hat{H}_{\mathrm{2DS}}=\hbar\sum_{j}\sum_{l=1}^2 q_j^{(l)}\ket{j}^{(l)}\bra{j}^{(l)}+\hbar\sum_{k} \omega_k\biggl[a_k^\dag a_k+|\alpha_k|^2\biggr]
    +\hbar\sum_{j,k}\sum_{l=1}^2\biggl[g_{j,k}^{(l)}\ket{j}^{(l)}\bra{j+1}^{(l)}(\alpha^*_k)+\textnormal{H.c.}\biggr],\\
         +\hbar\sum_{k} \biggl[a^\dag_k( \omega_k\alpha_k-i \dot{\alpha}_k)+a_k( \omega_k\alpha^*_k+i\dot{\alpha}^*_k)-i(\alpha^*_k-\alpha_k)\biggr]+\hbar\sum_{j,k}\sum_{l=1}^2\biggl[\chi_{j,k}^{(l)}\ket{j+1}^{(l)}\bra{j+1}^{(l)}a_k\alpha^*_{k}
         +\textnormal{H.c.}\biggr]
          \\+\hbar\sum_{j,k}\sum_{l=1}^2\left[\chi_{j,k}^{(l)}\ket{j+1}^{(l)}\bra{j+1}^{(l)}(a_k^\dag a_k+1)
         -\chi_{j,k}^{(l)}\ket{j}^{(l)}\bra{j}^{(l)}a_{k}^\dag a_k\right]
         -\hbar\sum_{j,k}\sum_{l=1}^2\biggl[\chi_{j,k}^{(l)}\ket{j}^{(l)}\bra{j}^{(l)}a_k^\dag \alpha_{k}+\textnormal{H.c.}\biggr]\\+
          \frac{\hbar}{2}\sum_{i,j,k}\left[\frac{{g}_{i,k}^{*(1)}{g}_{j,k}^{(2)}}{(q_{i,i+1}^{(1)}-\omega_k)}+\frac{{g}_{i,k}^{(1)}{g}_{j,k}^{*(2)}}{(q_{j,j+1}^{(2)}-\omega_k)}\right]\biggl(\ket{i}^{(2)}\bra{i+1}^{(2)}\ket{j+1}^{(1)}\bra{j}^{(1)}+\ket{i+1}^{(2)}\bra{i}^{(2)}\ket{j}^{(1)}\bra{j+1}^{(1)}\biggr)\\+\hbar
        \sum_{j,k}\sum_{l=1}^2\biggl[\big(B_{j,k}^{(l)} \big)^*\ket{j+1}^{(l)}\bra{j}^{(l)}( \omega_k\alpha_k-i \dot{\alpha}_k)+\textnormal{H.c.}\biggr],
     \end{multline}
\end{widetext}
where we have again dropped terms of $O\big(\big[B^{(l)}_{j,k}\big]^2\big)$ and smaller and those corresponding to two-photon transitions.

%% file: Open_Quantum_System.tex
\section{Open Quantum System Interpretation of Maxwell-Schr\"{o}dinger Methods}
\label{sec:Open-Quant}
To support our derivation of how to incorporate multi-qubit effects into Maxwell-Schr\"{o}dinger methods, we need to verify in what way the influence of the $\alpha_k$-dependent terms in (\ref{Eq:Disp-SW-BA}) are captured by the Maxwell-Schr\"{o}dinger method based on solving (\ref{eq:schro-wave4}) and (\ref{eq:phi-coupled-wave}) together. Unfortunately, a direct implementation of (\ref{Eq:Disp-SW-BA}) in a numerical simulation is not computationally tractable. To address this, in this appendix we develop a non-Markovian open quantum system simulation strategy that is computationally tractable for the more fundamental Hamiltonian given in (\ref{Sec2B:Eq:QH_full}). The theoretical formulation for this is given in Appendix \ref{subsec:open-quantum-formalism}, while a simple numerical discretization strategy for the model is presented in Appendix \ref{subsec:open-quantum-discretization}. Numerical results demonstrating the agreement between the Maxwell-Schr\"{o}dinger method and the open quantum system model developed here are presented in Appendix \ref{subsec:open-quantum-results}.

\subsection{Theoretical Model}
\label{subsec:open-quantum-formalism}

\begin{figure}[t]
    \centering
    \includegraphics[width=0.9\linewidth]{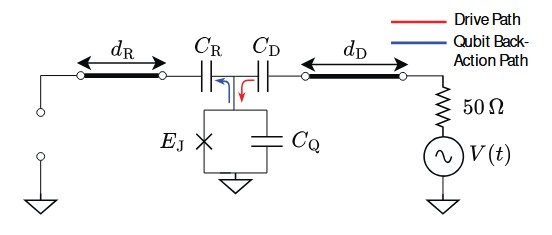}
    \caption{Simple cQED circuit with a single transmon capacitively connected through $C_\mathrm{D}$ and a transmission line of length $d_\mathrm{D}$ to a drive source. The transmon is also connected through $C_\mathrm{R}$ to a transmission line resonator of length $d_\mathrm{R}$ which is left open on one end. The arrows are provided to correspond to the settings used in our simulations in Section \ref{subsec:open-quantum-results} to focus our analysis only on the qubit back-action effect due to coupling with the resonator. }
    \label{fig:SingleCircuit}
\end{figure}

To make the quantum formulation more concrete, we will consider the specific case of a transmon capacitively coupled to an open-circuited transmission line resonator. We further assume that the device operates in the dispersive regime so that the qubit transition frequency is significantly detuned from the first resonant frequency of the transmission line resonator \cite{blais2021circuit}. Additionally, we consider that the qubit can also be directly driven by another voltage source capacitively coupled to the qubit through another transmission line. The overall circuit schematic is illustrated in Fig. \ref{fig:SingleCircuit}.

In principle, a complete description of this system should include many transmission line resonator modes to faithfully describe the spatial distributions that the resonator node flux can exhibit. This is especially relevant for considering qubit control scenarios because the frequency of the applied drive will typically occur at the significantly detuned value of the qubit transition frequency, leading to individual spatial mode patterns not accurately resolving the node flux distribution. Considering this, the Hamiltonian for the circuit of Fig. \ref{fig:SingleCircuit} can be given by
\begin{align}
    \label{Sec2B:Eq:H_totl}
    \hat{H}=\hat{H}_0+\hat{H}_I,
\end{align}
where
\begin{align}
    \label{Sec2B:Eq:H_O}
\hat{H}_{0}&=\hbar\sum_j  q_j\ket{j}\bra{j}+\hbar\sum_k \omega_k\hat{a}_k^\dag \hat{a}_k,\\
\label{Sec2B:Eq:H_I}
\hat{H}_{I}&= \hbar\bar{V}_\mathrm{D}(t)\hat{n} +  \hbar\sum_k g_k\hat{n}(\hat{a}_k+\hat{a}_k^\dag).
\end{align}
Here, $\omega_k$ is the frequency associated with the $k$-th mode of the resonator with corresponding annihilation (creation) operator $\hat{a}_k$ ($\hat{a}_k^\dag$). We note that the second term in (\ref{Sec2B:Eq:H_I}) can be shown to be equivalent to the interaction term in (\ref{Sec2B:Eq:QH_full}) by expressing the charge operator $\hat{n}$ in terms of the transmon eigenstates and taking a RWA. We also include in (\ref{Sec2B:Eq:H_I}) the term reflecting the effective voltage seen by the qubit due to the voltage source in angular frequency units as $\bar{V}_\mathrm{D}(t)=2e\beta_\mathrm{D} V(t)/\hbar$, where $\beta_\mathrm{D}=C_\mathrm{D}/C_\Sigma$ and $C_\Sigma=C_\mathrm{Q}+C_\mathrm{R}+C_\mathrm{D}$. Finally, the coupling between the qubit and $k$-th resonator mode is $g_k$, which is given by
\begin{align}
    \label{Eq:g_k}
    \hbar g_k=2e\beta_\mathrm{R}\sqrt{\frac{\hbar\omega_k}{d_\mathrm{R}C }} \cos\bigg( \frac{k \pi}{d_\mathrm{R}} z_0 \bigg).
\end{align}
In (\ref{Eq:g_k}), $\beta_\mathrm{R} = C_\mathrm{R}/C_\Sigma$ and the cosine accounts for the node flux spatial mode distribution in the open-circuited resonator shown in Fig. \ref{fig:SingleCircuit}. Here, $z_0$ is the location in the resonator that the qubit is coupled to, which for the case of Fig. \ref{fig:SingleCircuit} is $z_0 = d_\mathrm{R}$.



We can now formulate the desired open quantum system description of this model that will match the results from the Maxwell-Schr\"{o}dinger method. To do this, we first consider the dynamics in the interaction picture so that the time evolution of the full density matrix of the qubit-resonator system is given by
\begin{align}
    \label{eq:full_density_EoM}
    \frac{d\tilde{\rho}(t)}{dt}=-\frac{i}{\hbar}\biggl[\tilde{H}_I(t),\tilde{\rho}(t)\biggr],
\end{align}
where a tilde represents that a quantity is expressed in the interaction picture according to the free Hamiltonian $\hat{H}_0$. We will make the natural decomposition of our density matrix into the part associated with the qubit and another part associated with the resonator. More specifically, the reduced density matrix for the qubit will be given by $\tilde{\rho}_\mathrm{Q} = \mathrm{Tr}_\mathrm{R}\{ \tilde{\rho}\}$, where $\mathrm{Tr}_\mathrm{R}$ denotes a partial trace over all resonator states. Similarly, the reduced density matrix for the resonator will be given by $\tilde{\rho}_\mathrm{R} = \mathrm{Tr}_\mathrm{Q}\{ \tilde{\rho}\}$, where $\mathrm{Tr}_\mathrm{Q}$ is the partial trace over the qubit states.

To proceed with our derivation, we assume that the standard Born approximation holds \cite{breuer2002theory} so that we may write the full density matrix as a product state $\tilde{\rho}(t)\approx\tilde{\rho}_\mathrm{Q}(t)\otimes\tilde{\rho}_\mathrm{R}(t)$ for all $t$. However, we will retain the time dependence of $\tilde{\rho}_\mathrm{R}(t)$ to capture the non-Markovian effects of interest here. We also note that the Born approximation is a natural one for typical cQED devices that operate in the dispersive regime, as they are generally engineered to be weakly coupled to their transmission line environment to mitigate decoherence and crosstalk \cite{blais2021circuit}. Further, the Born approximation is necessary to recover the behavior of a Maxwell-Schr\"{o}dinger method, which due to the classical treatment of the transmission lines naturally cannot describe any entanglement between them and the qubit.

Considering these approximations, we can now derive an equation of motion for $\tilde{\rho}_\mathrm{Q}(t)$ by taking the partial trace of (\ref{eq:full_density_EoM}) over the resonator states. Doing this and removing the explicit time dependence of the operators for brevity, we arrive at
\begin{align}
    \label{eq:rho_q_EoM}
   \frac{d}{dt} \tilde{\rho}_\mathrm{Q}(t) =-i\bigg(  \bar{V}_\mathrm{R}(t)  +{\bar{V}_\mathrm{D}(t)} \bigg) [\tilde{n},\tilde{\rho}_\mathrm{Q}(t)]. 
\end{align}
where
\begin{align}
   \bar{V}_\mathrm{R}(t) = \sum_k g_k \mathrm{Tr}_\mathrm{R}\bigg\{ (\tilde{a}_k+\tilde{a}_k^\dag)\tilde{\rho}_\mathrm{R}(t)\bigg\}.
\end{align}
From (\ref{eq:rho_q_EoM}), we see that in this perspective the action of the resonator on the qubit in the form of $\bar{V}_\mathrm{R}(t)$ acts as a correction to the applied voltage felt by the qubit. 

However, to use this equation, we must have some way to approximate the behavior of $\tilde{\rho}_\mathrm{R}(t)$. While there are many different strategies that have been developed to model non-Markovian open quantum system dynamics \cite{breuer2002theory}, to recreate the effects of a Maxwell-Schr\"{o}dinger method we can follow a simplified approach that for the purposes of our validation in Section \ref{subsec:open-quantum-results} will remain computationally tractable. While it may seem natural in attempting to recreate a Maxwell-Schr\"{o}dinger method to assume $\tilde{\rho}_\mathrm{R}(t)$ is in a coherent state, we will see that this is unnecessary to be explicitly used in the numerical strategy given here. Instead, in analogy to (\ref{eq:rho_q_EoM}), we will formulate an equation of motion for $\tilde{\rho}_\mathrm{R}(t)$ by taking the partial trace of (\ref{eq:full_density_EoM}) over the qubit states to get
\begin{align}
    \label{eq:rho_r_EoM}
     \frac{d}{dt}\tilde{\rho}_\mathrm{R}(t) = -i \sum_k g_k \expval{n(t)} \biggl[\tilde{a}_k+\tilde{a}_k^\dag,\tilde{\rho}_\mathrm{R}(t)\biggr],
\end{align}
where
\begin{align}
    \expval{n(t)} = \mathrm{Tr}_\mathrm{Q} \bigg\{  \tilde{n} \tilde{\rho}_\mathrm{Q}(t)   \bigg\}.
\end{align}
We see that the qubit dynamics couples into (\ref{eq:rho_r_EoM}) through the expectation value of the charge operator, just as is done in the Maxwell-Schr\"{o}dinger equation given in (\ref{eq:phi-coupled-wave}). Correspondingly, we will follow a similar time evolution strategy to solve the coupled equations (\ref{eq:rho_q_EoM}) and (\ref{eq:rho_r_EoM}). 

\subsection{Numerical Discretization}
\label{subsec:open-quantum-discretization}
The simplest way to solve the coupled equations (\ref{eq:rho_q_EoM}) and (\ref{eq:rho_r_EoM}) is with a leap-frog time-marching process. To do this, we discretize the temporal derivatives with central finite differences as 
\begin{align}
    \label{eq:central_difference}
    \frac{df (t)}{dt}\bigg\rvert_{t=m\Delta t}\approx\frac{f^{(m+1)}-f^{(m-1)}}{2\Delta t},
\end{align}
where $\Delta t$ is the size of the time step of the time-marching process, $m$ is an integer, and we denote a quantity evaluated at the $m$-th time step with the shorthand notation $f(m\Delta t) = f^{(m)}$ \cite{jin2011theory}.

Using typical conventions for leap-frog time marching, we first discretize (\ref{eq:rho_q_EoM}) at integer time steps and (\ref{eq:rho_r_EoM}) at half-integer time steps. We then solve for the most time-advanced quantity in the respective equations to get
\begin{multline}
    \label{eq:rho_q_disc1}
    \tilde{\rho}_\mathrm{Q}^{(m+1)} = \tilde{\rho}_\mathrm{Q}^{(m-1)} -i2 \Delta t \bigg(  \bar{V}_\mathrm{R}^{(m)}  +{\bar{V}_\mathrm{D}^{(m)}} \bigg) \\ \times [\tilde{n}^{(m)},\tilde{\rho}_\mathrm{Q}^{(m)}], 
\end{multline}
\begin{multline}
    \label{eq:rho_r_disc1}
    \tilde{\rho}_\mathrm{R}^{(m+3/2)} = \tilde{\rho}_\mathrm{R}^{(m-1/2)} -i 2 \Delta t \sum_k g_k \expval{n^{(m+1/2)}} \\ \times \biggl[\tilde{a}_k^{(m+1/2)}+\tilde{a}_k^{\dag,(m+1/2)},\tilde{\rho}_\mathrm{R}^{(m+1/2)}\biggr].
\end{multline}
Turning to (\ref{eq:rho_q_disc1}), we have the issue that $\tilde{\rho}_\mathrm{R}$ (and correspondingly $\bar{V}_\mathrm{R}$) is sampled at half-integer time steps, and so is not known at time $t = m\Delta t$. The standard solution to this for finite difference methods is to compute the value $\bar{V}_\mathrm{R}^{(m)}$ as an average of the adjacent two half-integer time step values which are known, which also has the benefit of maintaining the same level of accuracy as the central finite difference approximation given in (\ref{eq:central_difference}) \cite{jin2011theory}. Doing this for (\ref{eq:rho_q_disc1}) and for the similar term in (\ref{eq:rho_r_disc1}), we arrive at
\begin{multline}
    \label{eq:rho_q_disc2}
    \tilde{\rho}_\mathrm{Q}^{(m+1)} = \tilde{\rho}_\mathrm{Q}^{(m-1)}  -i \Delta t \\ \times \bigg(  \bar{V}_\mathrm{R}^{(m+1/2)}+\bar{V}_\mathrm{R}^{(m-1/2)}  + 2 {\bar{V}_\mathrm{D}^{(m)}} \bigg)  [\tilde{n}^{(m)},\tilde{\rho}_\mathrm{Q}^{(m)}], 
\end{multline}
\begin{multline}
    \label{eq:rho_r_disc2}
    \tilde{\rho}_\mathrm{R}^{(m+3/2)} = \tilde{\rho}_\mathrm{R}^{(m-1/2)} \\ -i  \Delta t \sum_k g_k \bigg( \expval{n^{(m+1)}} + \expval{n^{(m-1)}} \bigg) \\ \times \biggl[\tilde{a}_k^{(m+1/2)}+\tilde{a}_k^{\dag,(m+1/2)},\tilde{\rho}_\mathrm{R}^{(m+1/2)}\biggr],
\end{multline}
which can be successfully marched forward in time together. This model will be used to quantitatively validate the Maxwell-Schr\"{o}dinger method for a wide range of drives applied to the qubit. 

\subsection{Numerical Results}
\label{subsec:open-quantum-results}

We now turn to simulating the circuit shown in Fig. \ref{fig:SingleCircuit} for representative values for the different circuit components. More specifically, the coupling capacitance between the qubit and the drive is $C_\mathrm{D}=0.1\;\mathrm{fF}$ and the total qubit capacitance is fixed at $C_\Sigma=67.95\;\mathrm{fF}$ so that we have a fixed $E_\mathrm{C}$ for all our simulations. The Josephson energy $E_\mathrm{J}$ is varied to alter the transmon's first transition frequency and we also change $C_\mathrm{R}$ (and consequently $C_\mathrm{Q}$) to illustrate the impact of these parameters on the qubit back-action. Further, to operate in the dispersive regime, we set our parameters so that the first mode of the transmission line resonator has a resonant frequency of $6.31\;\mathrm{GHz}$. This is achieved here by using a per-unit-length capacitance of $C=280\;\mathrm{pF/m}$, a per-unit-length inductance of $L=0.7\;\mu\mathrm{H/m}$, and a resonator length of $d_\mathrm{R}=5.66\;\mathrm{mm}$. The drive source for these simulations is a typical modulated Gaussian pulse given by
\begin{align}
    \label{sec3A:Eq:Voltage1Q}
    V(t)=V_{\mathrm{mag}}\sin(2\pi f_q [t-t_0])\exp\left(\frac{1}{2}\frac{(t-t_0)^2}{\sigma^2}\right),
\end{align}
where $V_{\mathrm{mag}}$ is the amplitude of the pulse, $t_{0}$ is a time offset, $\sigma$ is standard deviation of the Gaussian, and $f_q$ is the frequency of the pulse which is always kept resonant with the first transition frequency of the qubit.

As with the main text, such a circuit can be easily discretized and simulated using the Maxwell-Schr\"{o}dinger framework,but exactly replicating the impact of the various circuit connections in the open quantum system model that utilizes an eigenmode decomposition of the transmission line system can be laborious. Here, we make the same two simplifications within the Maxwell-Schr\"{o}dinger model to make the quantitative comparison with the open quantum system model easier. Namely, we only allow the applied voltage drive to affect the transmon and there is no residual leakage directly from the drive line to the resonator. Second, we only allow the back-action of the transmon to affect the resonator so that we can isolate the impact of this effect in comparing between the Maxwell-Schr\"{o}dinger and open quantum system models. These two simplifications are shown by the red and blue arrows in Fig. \ref{fig:SingleCircuit}.


As mentioned previously, correctly resolving the spatial distribution of the node flux in a resonator at a frequency that is significantly detuned from the resonator frequency requires including many modes of the resonator in the expansion in (\ref{Sec2B:Eq:H_I}). Truncations of this modal expansion can further suffer from a Gibbs-like phenomenon, causing the node flux distribution at the end point of the resonator that couples to the qubit to differ significantly from the real distribution. These issues can be circumvented for our current purposes in a simple way by considering a single mode of the resonator with a modified frequency-dependent coupling strength to the qubit to correct for the true standing-wave pattern in the resonator. For the system shown in Fig. \ref{fig:SingleCircuit} and for drive frequencies below the first fundamental frequency of the resonator, the node flux has a spatial distribution given approximately by $\cos(2\pi z / \lambda_q )$, where $\lambda_q$ is the wavelength at the drive frequency and $ 0 \leq z \leq d_\mathrm{R}$. For a typical qubit control scenario, this wavelength will be for the first qubit transition frequency $q_{0,1}=q_1-q_0$. Considering that we need to know the value of the standing-wave pattern at $z=d_\mathrm{R}$ (which is half of a wavelength at $\omega_1$), we can substitute into the spatial distribution and find the value will be $\cos(\pi   q_{0,1} / \omega_1 )$. Correspondingly, the modified frequency-dependent coupling strength that should be used in the single-mode approximate model causes (\ref{Eq:g_k}) to become
\begin{align}
    \label{Eq:g_Qutip}
    \hbar g=2e\beta_\mathrm{R}\sqrt{\frac{\hbar\omega_1}{d_\mathrm{R}C}} \cos\bigg( \frac{\pi q_{0,1}}{\omega_1} \bigg).
\end{align}
This approximation naturally only works effectively for reasonably narrowband drives, but this is still sufficient to test the agreement between the Maxwell-Schr\"{o}dinger and open quantum system models. 

\begin{figure}[t!]
    \centering
    \includegraphics[width=\linewidth]{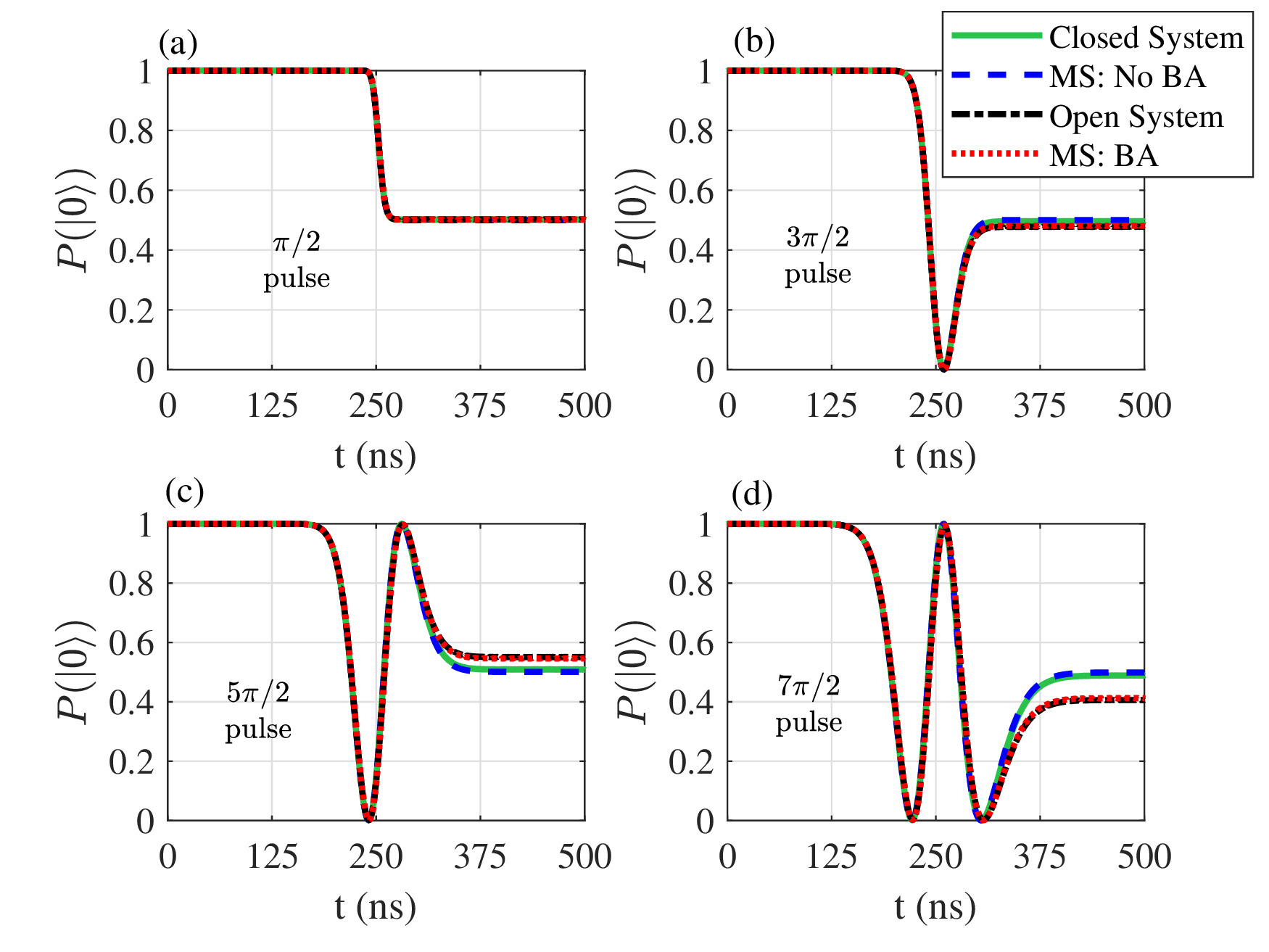}
    \caption{Occupation probability of the $\ket{0}$ state of a three-level transmon with first transition frequency $q_{0,1}/2\pi=4.6\;\mathrm{GHz}$ and $C_\mathrm{R}=6\;\mathrm{fF}$. We compare the state dynamics across four different methods. To illustrate the state dynamics when back-action is absent, we use the Maxwell–Schr\"{o}dinger approach with no back-action (MS: No BA) and the direct time-evolution of the closed system Hamiltonian (Closed System). The methods accounting for the back-action are the full Maxwell–Schr\"{o}dinger (MS: BA) and the open quantum system (Open System) methods. The dynamics are reported for four different driven Rabi pulse areas: (a) $\pi/2$, (b) $3\pi/2$, (c) $5\pi/2$, and (d) $7\pi/2$.}
    \label{fig:Q=4.6,C=6}
\end{figure}

To begin these tests, we use a qubit with $q_{0,1}/2\pi=4.6\;\mathrm{GHz}$ and set the coupling capacitance to the resonator to $C_\mathrm{R}=6\;\mathrm{fF}$. For all tests, we consider a transmon with three energy levels included in the discretization. We then compare the results between four different simulation methods for a variety of driven Rabi pulses in Fig. \ref{fig:Q=4.6,C=6} to demonstrate the impact of the qubit back-action on the qubit dynamics. The first two models focus on simulations that do not account for the qubit back-action, which serve as a baseline for conventional simulations of qubit dynamics. The first is a standard closed quantum system model with Hamiltonian given in (\ref{Sec2B:Eq:H_totl}) using the single-mode approximation discussed in this section. These closed quantum system dynamics are then solved using QuTiP \cite{johansson2012qutip}. The second model ignoring qubit back-action is our Maxwell-Schr\"{o}dinger model with the back-action artificially turned off by removing the term on the right-hand side of (\ref{eq:phi-coupled-wave}). The strong agreement across different driven Rabi pulse types shows that the definitions of the voltage pulses between the Maxwell-Schr\"{o}dinger and QuTiP simulations are calibrated correctly with each other.

The next two simulation types account for the qubit back-action, and correspond to using the full Maxwell-Schr\"{o}dinger model and the open quantum system model. The open quantum system model is solved using the time-stepping approach given in (\ref{eq:rho_q_disc2}) and (\ref{eq:rho_r_disc2}) and also uses the same single-mode approximation as the closed quantum system model. Using the same applied voltage definitions as were used for the simulations that ignored back-action, we see in Fig. \ref{fig:Q=4.6,C=6} that the impact of this back-action can become substantial for longer duration pulses. We also see that there is excellent quantitative agreement between the Maxwell-Schr\"{o}dinger and open quantum system models, demonstrating the validity of our proposed interpretation of Maxwell-Schr\"{o}dinger methods as capturing a class of non-Markovian open quantum system effects.

\begin{figure}[t!]
    \centering
    \includegraphics[width=\linewidth]{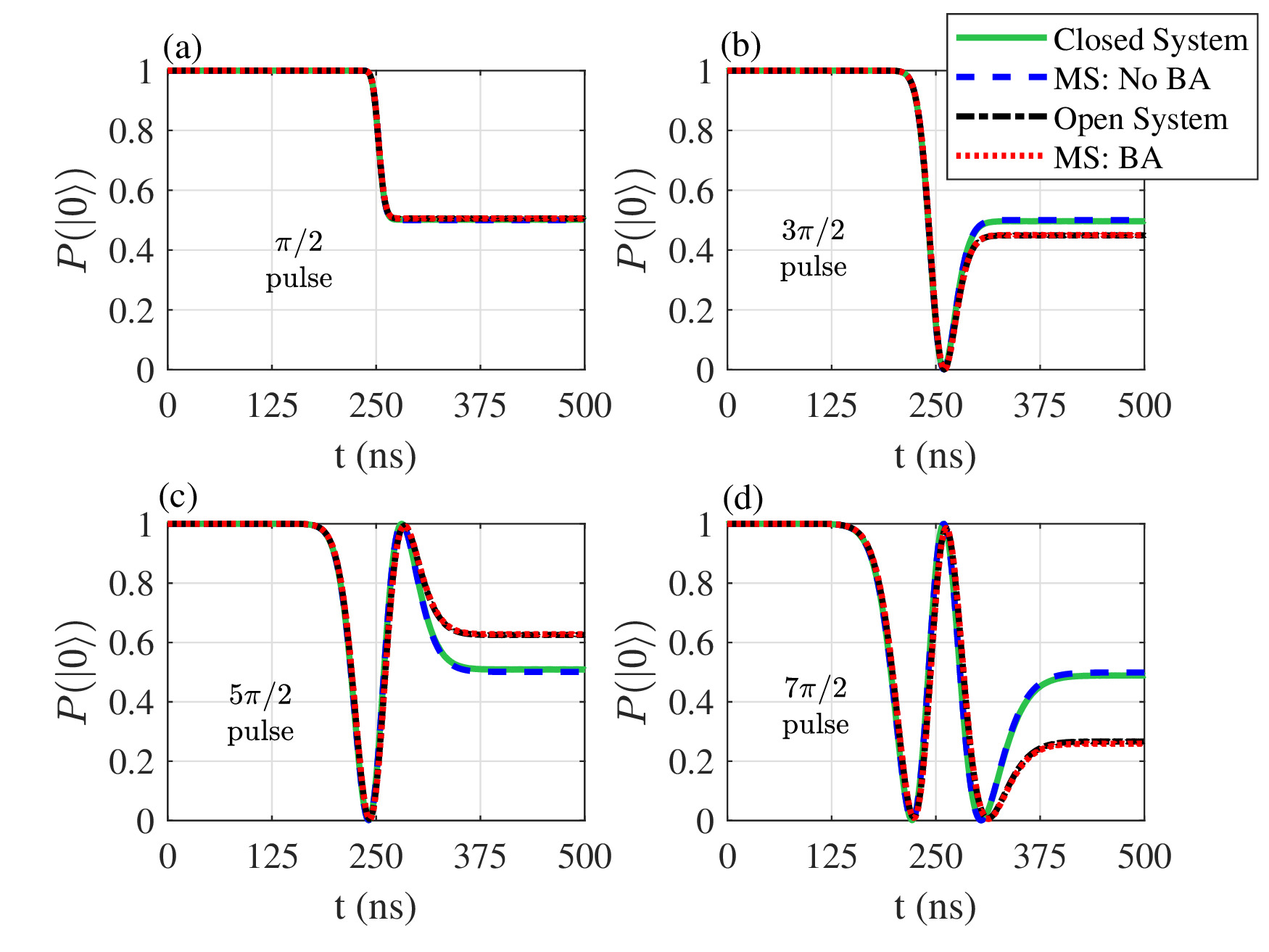}
    \caption{The simulations performed in Fig. \ref{fig:Q=4.6,C=6} are repeated, but with the coupling capacitance increased to $C_\mathrm{R}=8\;\mathrm{fF}$ and the first transition frequency kept at $q_{0,1}/2\pi=4.6\;\mathrm{GHz}$.}
    \label{fig:Q=4.6,C=8}
\end{figure}
\begin{figure}[t!]
    \centering
    \includegraphics[width=\linewidth]{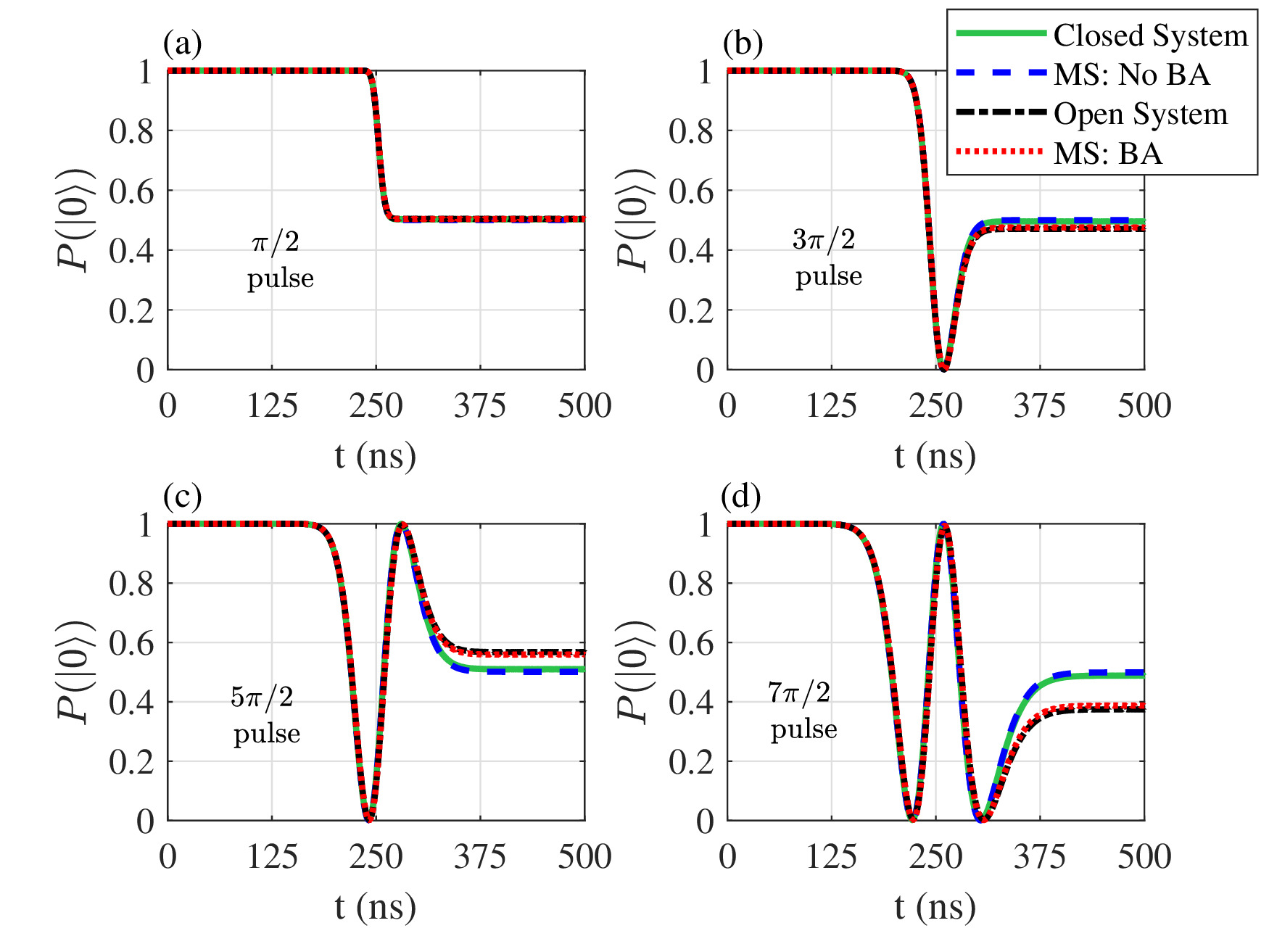}
    \caption{The simulations performed in Fig. \ref{fig:Q=4.6,C=6} are repeated, but with the first transition frequency increased to $q_{0,1}/2\pi=4.7\;\mathrm{GHz}$ and the coupling capacitance kept at $C_\mathrm{R}=6\;\mathrm{fF}$.}
    \label{fig:Q=4.7,C=6}
\end{figure}

To further test these conclusions, we modify some of the key parameters impacting the qubit back-action and repeat the simulations for various driven Rabi pulse types. In Fig. \ref{fig:Q=4.6,C=8}, we keep the qubit frequency at $q_{0,1}/2\pi=4.6\;\mathrm{GHz}$ and change the coupling capacitance to the resonator to $C_\mathrm{R}=8\;\mathrm{fF}$. The increased coupling capacitance increases the amount of back-action and subsequently the difference between the idealized models that ignore back-action and the models that incorporate this effect. Next, in Fig. \ref{fig:Q=4.7,C=6}, we repeat these tests again but for a qubit frequency of $q_{0,1}/2\pi=4.7\;\mathrm{GHz}$ and a coupling capacitance of $C_\mathrm{R}=6\;\mathrm{fF}$. With the decrease in detuning between the qubit and resonator frequency, we would expect a greater impact of the back-action, which is correspondingly observed in Fig. \ref{fig:Q=4.7,C=6}.

\begin{figure}[t!]
    \centering
    \includegraphics[width=0.95\linewidth]{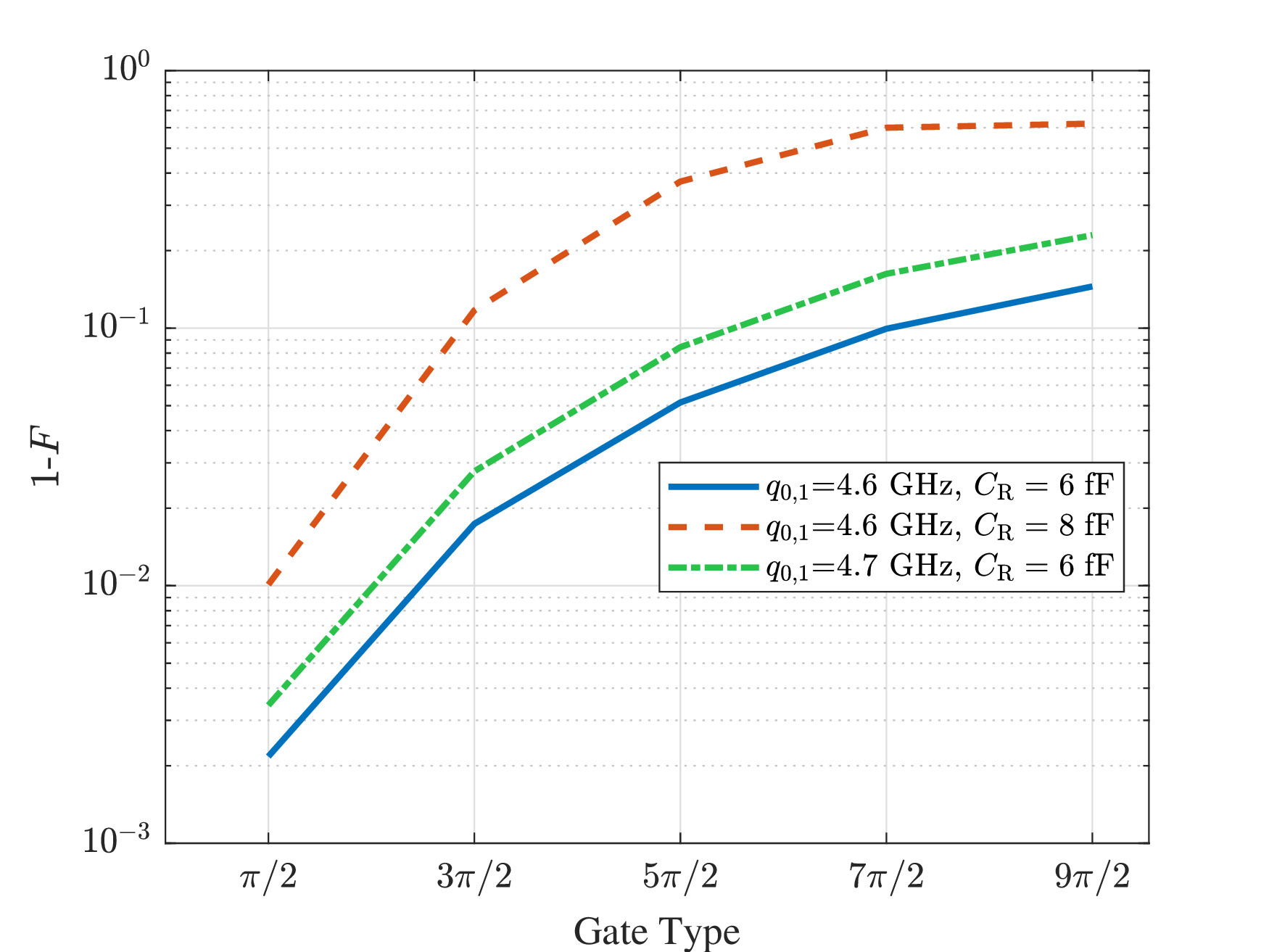}
    \caption{Plotting the infidelity ($1-F$) against Rabi pulse types for the simulations performed in Figs. \ref{fig:Q=4.6,C=6} to \ref{fig:Q=4.7,C=6}. The infidelity is calculated between the Maxwell-Schr\"{o}dinger simulations with and without back-action using (\ref{Eq:fidelity}).}
    \label{fig:fidelity}
\end{figure}

To quantitatively demonstrate the difference between the gate predicted by the Maxwell-Schr\"{o}dinger framework without back-action and with back-action, we plot the fidelities corresponding to these methods for the simulations run in Figs. \ref{fig:Q=4.6,C=6} to \ref{fig:Q=4.7,C=6} in Fig. \ref{fig:fidelity}. To calculate this quantity, we repeat our simulations for starting the qubit in the $\ket{0}$ and $\ket{1}$ states. We use the results from the Maxwell-Schr\"{o}dinger simulation with no back-action to construct the unitary for the ideal simulation, denoted as $\hat{U}$. We also perform these simulations with the Maxwell-Schr\"{o}dinger method with back-action and record the final states as $\ket{\psi_0}$, and $\ket{\psi_1}$, where the subscript denotes what the starting state was for the simulation. Using these quantities, the fidelity can be computed as \cite{nielsen2002simple}
\begin{align}
    \label{Eq:fidelity}
    F=\frac{1}{2}\sum_{j={0,1}}\bra{j}\hat{U}^\dag\ket{\psi_j}\bra{\psi_j}\hat{U}\ket{j}.
\end{align}
We see that the impact of the back-action increases as the pulse duration increases. Further, we see that the severity of the back-action effect can greatly depend on the different parameters of the circuit and qubit design.